\documentclass[11pt,reqno]{amsart}
\usepackage{amsaddr}
\usepackage{setspace}
\usepackage{graphicx}
\usepackage{dsfont}
\usepackage{algorithm}
\usepackage{algpseudocode}
\usepackage{enumerate}
\usepackage{mathtools}
\usepackage{hyperref}
\usepackage{amsmath}
\usepackage{amssymb}
\usepackage{tikz}
\usepackage{hyperref}

\usepackage[numbers]{natbib}
\setcitestyle{aysep={}}
\usepackage[babel,english=british]{csquotes}

\usepackage{xcolor}

\newcommand{\norm}[1]{\left\lVert#1\right\rVert}
\newcommand{\IR}{\mathbb{R}}
\newcommand{\IN}{\mathbb{N}}

\newcommand{\IC}{\mathbb{C}}

\usepackage{amsthm}
\newtheorem{theorem}{Theorem}[section]

\newtheorem{corollary}[theorem]{Corollary}

\newtheorem{lemma}[theorem]{Lemma}

\theoremstyle{definition}

\newtheorem{remark}[theorem]{Remark}

\numberwithin{equation}{section}

\DeclareMathOperator*{\argmax}{arg\,max}

\renewenvironment{proof}{{\bfseries Proof:}}{\hfill $\square$}

\begin{document}

\title{Quantum Learning Boolean Linear Functions w.r.t. Product Distributions}
\author{Matthias C. Caro}   
\address{Technical University of Munich, Germany, Department of Mathematics}
\email{\href{mailto:caro@ma.tum.de}{caro@ma.tum.de}, \emph{ORCID:} \href{https://orcid.org/0000-0001-9009-2372}{0000-0001-9009-2372}}
\date{\today}

\begin{abstract}
The problem of learning Boolean linear functions from quantum examples w.r.t.~the uniform distribution can be solved on a quantum computer using the Bernstein-Vazirani algorithm \cite{Bernstein.1993}. A similar strategy can be applied in the case of noisy quantum training data, as was observed in \cite{Grilo.20180410}. However, extensions of these learning algorithms beyond the uniform distribution have not yet been studied. We employ the biased quantum Fourier transform introduced in \cite{Kanade.2019} to develop efficient quantum algorithms for learning Boolean linear functions on $n$ bits from quantum examples w.r.t.~a biased product distribution. Our first procedure is applicable to any (except full) bias and requires $\mathcal{O}(\ln (n))$ quantum examples. The number of quantum examples used by our second algorithm is independent of $n$, but the strategy is applicable only for small bias. Moreover, we show that the second procedure is stable w.r.t.~noisy training data and w.r.t.~faulty quantum gates. This also enables us to solve a version of the learning problem in which the underlying distribution is not known in advance. Finally, we prove lower bounds on the classical and quantum sample complexities of the learning problem. Whereas classically, $\Omega (n)$ examples are necessary independently of the bias, we are able to establish a quantum sample complexity lower bound of $\Omega (\ln (n))$ only under an assumption of large bias. Nevertheless, this allows for a discussion of the performance of our suggested learning algorithms w.r.t.~sample complexity. With our analysis we contribute to a more quantitative understanding of the power and limitations of quantum training data for learning classical functions.
\end{abstract}

\maketitle

\section{Introduction}
The origins of the fields of machine learning as well as quantum information and computation both lie in the $1980$s. The arguably most influential learning model, namely the PAC (``probably approximately correct'') model, was introduced by Valiant in $1984$ \cite{Valiant.1984} with which the problem of learning was given a rigorous mathematical framework. Around the same time, Benioff \cite{Benioff.1980} and Feynman presented the idea of quantum computers \cite{Feynman.1985} to the public and thus gave the starting signal for important innovations at the intersection of computer science, information theory and quantum theory. Both learning theory and quantum computation promise new realms of computation in which tasks that seem insurmountable from the perspective of classical computation become feasible. The first has already proved its practical worth and is indispensable for modern-world big data applications, the latter is not yet as practically relevant but much work is invested to make the promises of quantum computation a reality. The interested reader is referred to \cite{ShalevShwartz.2014} and \cite{Nielsen.2010} for an introduction to statistical learning and quantum computation and information, respectively.\\

Considering the increasing importance of machine learning and quantum computation, attempting a merger of the two seems a natural step to take and the first step in this direction was taken already in \cite{Bshouty.1998}. The field of quantum learning has received growing attention over the last few years and by now some settings are known in which quantum training data and the ability to perform quantum computation can be advantageous for learning problems from an information-theoretic as well as from a computational perspective, in particular for learning problems with fixed underlying distribution (see e.g.~\cite{Arunachalam.2017} for an overview). It was, however, shown in \cite{Arunachalam.2018} that no such information-theoretic advantage can be obtained in the (distribution-independent) quantum PAC model (based on \cite{Bshouty.1998}) compared to the classical PAC model (introduced in \cite{Valiant.1984}).\\

One of the early examples of the aptness of quantum computation for learning problems is the task of learning Boolean linear functions w.r.t.~the uniform distribution via the Bernstein-Vazirani algorithm presented in \cite{Bernstein.1993}. Whereas this task of identifying an unknown $n$-bit string classically requires a number of examples growing (at least) linearly with $n$, a bound on the sufficient number of copies of the quantum example state independent of $n$ can be established. This approach was taken up in \cite{Grilo.20180410} where it is shown that, essentially, the Bernstein-Vazirani-based learning method is also viable if the training data is noisy. However, also this analysis is restricted to quantum training data arising from the uniform distribution. The same limiting assumption was also made in \cite{Bshouty.1998} for learning Disjunctive Normal Forms and in this context an extension to product distributions was achieved in \cite{Kanade.2019}.\\

Hence, a next direction to go is building up on the reasoning of \cite{Kanade.2019} to extend the applicability of quantum learning procedures for linear functions to more general distributions. The analysis hereby differs from the one for DNFs because no concentration results for the biased Fourier spectrum of a linear function are available. Moreover, whereas many studies of specific quantum learning tasks focus on providing explicit learning procedures yielding a better performance than known classical algorithms, we complement our learning algorithms with lower bounds on the size of the training data for a comparison to the best classical procedure and for a discussion of optimality among possible quantum strategies.

\subsection{Overview over the Results}
The task of learning linear functions has already served as a toy model for quantum speed-ups in the early days of quantum computing. We describe possible generalizations of known results in different scenarios. First, in Theorem \ref{ThmAmpBiasBV1} we exhibit a Fourier-sampling-based algorithm which learns Boolean linear functions on $n$ inputs from $\mathcal{O}(\ln(n))$ quantum examples arising from a $c$-bounded product distribution $D_\mu$. (Classically, it is known that $\Omega (n)$ examples are required.) Moreover, for a bias vector $\mu$ satisfying $|\mu_i|\leq\mathcal{O}\left(\frac{1}{\sqrt{n}} \right)$ for all $i$, this can be reduced to $\mathcal{O}(1)$ quantum examples (Theorem \ref{ThmAmpBiasBV2}). We also show that this reduction to a constant number of quantum examples is not possible for arbitrary product distributions by giving quantum sample complexity lower bounds in Theorem \ref{ThmQuSampCompLowerBoundV2}.\\
In Theorem \ref{ThmAmpBiasBVNoisy}, we exhibit a noise bound for quantum examples arising from a product distribution $D_\mu$ with $|\mu_i|\leq\mathcal{O}\left(\frac{1}{\sqrt{n}} \right)$ for all $i$ but corrupted by noise which guarantees that $\mathcal{O}(1)$ quantum examples still suffice for learning. Under milder assumptions on the noise, a $\mathcal{O}(\ln (n))$ upper bound on the sample complexity is given. Similarly, faulty quantum gates can be tolerated in our learning algorithm. Based on this observation, we construct a quantum learning algorithm without prior knowledge of the underlying distribution which requires $\mathcal{O}(n^2)$ quantum examples by first estimating the bias vector classically (Corollary \ref{CrlQuLearnLinUnknownDistr}).

\subsection{Related Work}
The (classical) problem of learning linear functions from randomly drawn examples in the presence of noise was studied in \cite{Blum.2003} (over the field $\mathbb{F}_2$) as well as in \cite{Regev.2009} (over a field $\mathbb{F}_q$ for $q$ prime). The latter of these two works also established the relevance of this learning problem for cryptography by connecting it to certain lattice problems. A different model for learning linear functions is studied in \cite{Ivanyos.2018}, where the training data is not assumed to be noisy but instead only partial information about the function values is revealed.\\
The quantum PAC model was introduced in \cite{Bshouty.1998}, where it was employed for learning DNF formulae w.r.t.~the uniform distribution using a quantum example oracle. This was extended to product distributions by \cite{Kanade.2019}. On the basis of this notion of quantum examples, the known Bernstein-Vazirani algorithm \cite{Bernstein.1993} can be reinterpreted as giving rise to a quantum learning algorithm for linear functions. This interpretation is explicitly given and further elaborated upon for the case of noisy training data in \cite{Cross.2015} (for $q$=2) and in \cite{Grilo.20180410} (for general primes $q$). \cite{Cross.2015} established that, whereas the learning parity problem without noise is feasible both for classical and quantum computation, the learning parity with noise problem is widely believed to be classically intractable but remains feasible for quantum computers, where the runtime depends only logarithmically on the number of qubits. This quantum advantage for noisy systems was demonstrated experimentally in \cite{Riste.2017}. \cite{Grilo.20180410} extends this analysis to general fields and a broader class of noise models and obtains that also for that scenario, learning linear functions from noisy data is feasible for quantum computers, however, their runtime bound is polynomial in the number of subsystems.\\
In \cite{Atc.2007}, the class of juntas is found to also allow for efficient quantum learning. The framework of Fourier-based quantum exact learning is shown to be efficiently applicable more generally also to Fourier-sparse functions in \cite{Arunachalam.20190307}. Limitations of the power of quantum computation for learning have been studied in a series of papers culminating in \cite{Arunachalam.2018} and, more recently also in \cite{Arunachalam.20190307b}. The former work shows that without prior restrictions on the underlying probability distribution, quantum examples are not more powerful than classical examples. The latter work demonstrates that, assuming quantum hardness of the learning with errors problem from classical examples, the class of shallow circuits is hard to learn from quantum examples.\\
Aside from the task of learning from examples, also the problem of learning from membership queries, both classical and quantum, is well studied. For instance, \cite{Servedio.2004} established a polynomial relation between the number of required quantum versus required classical queries, which was recently improved upon in \cite{Arunachalam.20190307}. Also, \cite{Montanaro.2012} uses quantum membership queries for learning multilinear polynomials more efficiently than is classically possible. 

\subsection{Structure of the Paper}
The paper is structured in the following way. In section \ref{SctPrelims} we introduce the well-known notions from classical learning, quantum computation and Boolean Fourier analysis required for our purposes as well as the prototypic learning algorithm which motivates our procedures. Section \ref{SctLearningProblem} consists of a description of the learning task to be considered. This is followed by a generalization of the Bernstein-Vazirani algorithm to product distributions in section \ref{SctGenBV}. In the next section, this is used to develop two quantum algorithms for solving our problem. (Appendix \ref{SctNoiseStability} contains a stability analysis of the second of the two procedures w.r.t.~noise in training data and computation.) In section \ref{SctSampCompLowerBound} we establish sample complexity lower bounds complementing the upper bounds implied by the algorithms of section \ref{SctQuantSampCompUpperBounds}. Finally, we conclude with some open questions and the references.

\section{Preliminaries}\label{SctPrelims}

\subsection{Basics of Quantum Information and Computation}
We first define some of the fundamental objects of quantum information theory, albeit restricted to those required in our discussion. For the purpose of our presentation, we will consider a pure $n$-qubit quantum state to be represented by a state vector $|\psi\rangle\in\IC^{2^n}$ (in Dirac notation). Such a state encodes measurement probabilities in the following way: If $\lbrace |b_i\rangle\rbrace_{i=1}^{2^n}$ is an orthonormal basis of $\IC^{2^n}$, then there corresponds a measurement to this basis and the probability of observing outcome $i$ for a system in state $|\psi\rangle$ is given by $|\langle b_i|\psi\rangle|^2$. Finally, when considering multiple subsystems we will denote the composite state by the tensor product, i.e. if the first system is in state $|\psi\rangle$ and the second in state $|\phi\rangle$, the composite system is in state $|\psi,\varphi\rangle:=|\psi\rangle\otimes |\phi\rangle$.\\

Quantum computation now consists in evolution of quantum states. Performing a computational step on an $n$-qubit state corresponds to applying an $2^n\times 2^n$ unitary transformation to the current quantum state. (The most relevant example of such unitary gates in our context will be the (biased) quantum Fourier transform discussed in more detail in subsection $2.4$.) As the outcome of a quantum computation is supposed to be classical, as final step of our computation we perform a measurement such that the final output will be a sample from the corresponding measurement statistics.\\

We will also use some standard notions from (quantum) information theory. For example, we denote the Shannon entropy of a random variable $X$ by $H(X)$, the conditional entropy of a random variable $X$ given $Y$ as $H(X|Y)$ and the mutual information between random variables $X$ and $Y$ as $I(X:Y)$. Similarly, the von Neumann entropy of a quantum state $\rho$ will be denoted as $S(\rho)$ and the mutual information for a bipartite quantum state $\rho_{AB}$ as $I(\rho_{AB})=I(A:B)$. Standard results on these quantities which will enter our discussion can e.g.~be found in \cite{Nielsen.2010}.

\subsection{Basics of Learning Theory}
Next we describe the model of exact learning. In classical exact learning for an input space $\mathcal{X}$, a target space $\{ 0,1\}$, and a concept class $\mathcal{F}\subset\{ 0,1\}^\mathcal{X}$, a learning algorithm receives as input labelled training data $\lbrace (x_i,f(x_i))\rbrace_{i=1}^m$ for some (to the learner) unknown $f\in\mathcal{F}$, where the $x_i$ are drawn independently according to some probability distribution $D$ on $\mathcal{X}$ which is known to the learner. The goal of the learner is to exactly reproduce the unknown function $f$ from such training examples with high success probability.\\
We can formalize this as follows: We call a concept class $\mathcal{F}$ exactly learnable if there exists a learning algorithm $\mathcal{A}$ and a map $m_\mathcal{F}:(0,1)\to\IN$ s.t.~for every $D\in\textrm{Prob}(X)$ (where $\textrm{Prob}(X)$ is the set of all probability measures on $X$), $f\in\mathcal{F}$  and $\delta\in (0,1)$, running $\mathcal{A}$ on training data of size $m\geq m_\mathcal{F}(\delta)$ drawn according to $D$ and $f$ with probability $\geq 1-\delta$ (w.r.t.~the choice of training data) yields a hypothesis $h$ s.t. $h(x)=f(x)$ for all $x\in\mathcal{X}$. The smallest such map $m_\mathcal{F}$ is called sample complexity of exactly learning $\mathcal{F}$.\\

Note that this definition of learning captures the information-theoretic challenge of the learning problem in the sample complexity, but it does not refer to the computational complexity of learning. The focus on sample complexity is typical in statistical learning theory. Hence, also our results will be formulated in terms of sample complexity bounds. As we give explicit algorithms, these results directly imply bounds on the computational complexity, however, we will not discuss them in any detail.\\
Note also that the exact learning model differs from the well-known PAC (``probably approximately correct''), introduced by \cite{Valiant.1984}, in two ways. First, whereas the PAC model only requires to approximate the unknown function with high probability, we require to reproduce it exactly, in other words, we set the accuracy in PAC learning to $0$. Second, whereas in the PAC scenario the learner does not know the underlying distribution, we assume it to be fixed and known in advance. A short discussion on how to relax this restriction can be found in subsection \ref{SbSctUnknownDistr}.\\

The quantum exact learning model differs from the classical model in the form of the training data and the allowed form of computation. Namely, in quantum exact learning, the training data consists of $m$ copies of the quantum example state $|\psi_f\rangle=\sum\limits_{x\in\mathcal{X}} \sqrt{D(x)}|x,f(x)\rangle$, and this training data is processed by quantum computational steps. With this small change the above definition of exact learnability and sample complexity now carry over analogously.\\

We conclude this introduction with a concentration result that has proven to be useful throughout learning theory.
\begin{lemma}\emph{(Hoeffding's Inequality \cite{Hoeffding.1963}, compare also Theorem $2.2.6$ in \cite{Vershynin.2018})}\label{LmmHoeffding}\\
Let $Z_1,...,Z_n$ be real-valued independent random variables taking values in closed and bounded intervals $[a_i,b_i]$, respectively. Then for every $\varepsilon >0$
\begin{align*}
\mathbb{P}\Big[\sum\limits_{i=1}^n Z_i-\mathbb{E}[Z_i] \geq\varepsilon\Big]\leq\exp\Big(-\frac{2\varepsilon^2}{\sum_{i=1}^n (a_i-b_i)^2}\Big).
\end{align*}
This directly implies (after replacing $Z_i$ with $-Z_i$) that
\begin{align*}
\mathbb{P}\Big[\Big\lvert\sum\limits_{i=1}^n Z_i-\mathbb{E}[Z_i]\Big\rvert\geq\varepsilon\Big]\leq 2\exp\Big(-\frac{2\varepsilon^2}{\sum_{i=1}^n (a_i-b_i)^2}\Big).
\end{align*}
\end{lemma}

\subsection{$\mu$-biased Fourier Analysis of Boolean Functions}
We now give the basic ingredients of $\mu$-biased Fourier analysis over the Boolean cube $\lbrace -1,1\rbrace^n$. For more details, the reader is referred to \cite{ODonnell.2014}.\\
For a bias vector $\mu\in [-1,1]^{n}$, define the $\mu$-biased product distribution $D_{\mu}$ on $\lbrace-1,1\rbrace^{n}$ via 
\begin{align*}
D_{\mu}(x):= & \left(\prod_{i:x_{i}=1}\frac{1+\mu_{i}}{2}\right)\left(\prod_{i:x_{i}=-1}\frac{1-\mu_{i}}{2}\right)=\prod_{1\leq i\leq n}\frac{1+x_{i}\mu_{i}}{2},\quad x\in\lbrace-1,1\rbrace^{n}.
\end{align*}
Thus, a positive $\mu_i$ tells us that at the $i^{th}$ position the distribution is biased towards $+1$, a negative $\mu_i$ tells us that at the $i^{th}$ position the distribution is biased towards $-1$. For $\mu = 0\ldots 0$ we simply obtain the uniform distribution on $\lbrace-1,1\rbrace^{n}$. The absolute value of $\mu_i$ quantifies the strength of the bias in the $i^{th}$ component. We call $D_{\mu}$ $c$-bounded, for $c\in(0,1]$, if $\mu\in[-1+c,1-c]^{n}.$ Assuming the underlying product distribution to be $c$-bounded thus corresponds to assuming that the bias is not arbitrarily strong. Hence, we will in the following express notions of ``small'' or ``large'' bias either in terms of the bias vector $\mu$ or in terms of the $c$-boundedness constant.\\
For Fourier analysis we now need an orthonormal basis for the function space $\IR^{\lbrace -1,1\rbrace^n}$ w.r.t.~the inner product $\langle .,. \rangle_\mu$ defined by $$\langle f,g\rangle_\mu = \mathbb{E}_{D_\mu}[fg]=\sum\limits_{x\in\{-1,1\}^n} f(x)g(x)D_\mu(x).$$ One can show (using the product structure to reduce to the case $n=1$) that such an orthonormal basis is given by $\lbrace \phi_{\mu,j}\rbrace_{j\in\lbrace 0,1\rbrace^n}$ with $\phi_{\mu,j}(x)=\prod\limits_{i:j_i=1}\frac{x_i-\mu_i}{\sqrt{1-\mu_i^2}}$. For a function $f:\lbrace -1,1\rbrace^n \to\lbrace -1,1\rbrace$ this now gives a representation $f(x)=\sum\limits_{j\in\lbrace 0,1\rbrace^n} \hat{f}_\mu (j) \phi_{\mu,j}(x)$ with $\hat{f}_\mu (j):=\langle f,\phi_{\mu,j} \rangle_\mu$. For $\mu = 0\ldots 0$ we recover the well-known orthonormal basis consisting of $\chi_j (x) = (-1)^{j\cdot x}$ from standard Fourier analysis over the Boolean cube.

\subsection{$\mu$-biased Quantum Fourier Sampling}\label{SbSctQuFourierSampling}
We now turn to the description of the quantum algorithm for $\mu$-biased quantum Fourier sampling which constitutes the basic ingredient of our learning algorithms and which, to our knowledge, was first presented in \cite{Kanade.2019}. There the authors demonstrate that the $\mu$-biased Fourier transform for a $c$-bounded $D_\mu$ with $c\in(0,1]$ can be implemented on a quantum computer as the $n$-qubit $\mu$-biased quantum Fourier transform: For $x\in\lbrace -1,1\rbrace^n,$
\begin{align*}
H_\mu^n |x\rangle = H_\mu\otimes\ldots\otimes H_\mu |x_1,\ldots,x_n\rangle = \sum\limits_{j\in\lbrace 0,1\rbrace^n} \sqrt{D_\mu(x)} \phi_{\mu,j}(x)|j\rangle.
\end{align*}
In the same way as the unbiased quantum Fourier transform can be used for quantum Fourier sampling, this $\mu$-biased version now yield a procedure to sample from the $\mu$-biased Fourier spectrum of a function using a quantum computer. We describe the corresponding procedure in Algorithm \ref{AlgBiasQFS}.

\begin{algorithm}
\caption{$\mu$-biased Quantum Fourier Sampling}\label{AlgBiasQFS}
\begin{flushleft}
\hspace*{\algorithmicindent} \textbf{Input}: $|\psi_{f}\rangle = \sum\limits_{x\in\lbrace -1,1\rbrace^n}\sqrt{D_\mu (x)}|x,f(x)\rangle$ for a function $f:\lbrace -1,1\rbrace^{n}\to\lbrace 0,1\rbrace$\\
\hspace*{\algorithmicindent} \textbf{Output}: $j\in\lbrace0,1\rbrace^{n}$ with probability $\left(\hat{g}_{\mu}(j)\right)^{2}$, where the function $g:\lbrace -1,1\rbrace^{n}\to\lbrace -1,1\rbrace$ is defined as $g(x) = (-1)^{f(x)}$.\\
\hspace*{\algorithmicindent} \textbf{Success Probability}: $\frac{1}{2}$ 
\end{flushleft}
\begin{algorithmic}[1]
\State Perform the $\mu$-biased QFT $H_\mu$ on the first $n$ qubits, obtain the state $(H_\mu \otimes \mathds{1})|\psi_{f}\rangle$.
\State Perform a Hadamard gate on the last qubit, obtain the state $(H_\mu \otimes H)|\psi_{f}\rangle$.
\State Measure each qubit in the computational basis and observe outcome $j = j_1\ldots j_{n+1}$.
\If{$j_{n+1} = 0$}\Comment{This corresponds to a failure of the sampling algorithm.}
	\State Output $o\gets\perp$ and end computation.
\ElsIf{$j_{n+1} = 1$}\Comment{This corresponds to a success of the sampling algorithm.}
	\State Output $o\gets j_1\ldots j_n$ and end computation.
\EndIf
\end{algorithmic}
\end{algorithm}

One can show that this algorithm indeed works as claimed by by analysing the transformation of the quantum state throughout the steps algorithm and making use of the orthonormality of the basis. This is the content of the following

\begin{lemma}\emph{(Lemma $3$ in \cite{Kanade.2019})}\label{LmmBiasedQFourierSampling}\\
Denote $g:\lbrace -1,1\rbrace^{n}\to\lbrace -1,1\rbrace$, $g(x) = (-1)^{f(x)}$. Then with probability $\frac{\left(\hat{g}_{\mu}(j)\right)^{2}}{2}$, Algorithm \ref{AlgBiasQFS} outputs the string $j\in\lbrace0,1\rbrace^{n}.$
\end{lemma}
\begin{proof}
The proof can be found in \cite{Kanade.2019}, we reproduce it in Appendix \ref{SctProofs}.
\end{proof}

This result allows us to generalize results based on quantum Fourier sampling w.r.t.~the uniform distribution. In particular, we will apply it to obtain a generalization of the Bernstein-Vazirani algorithm.

\subsection{The Pretty Good Measurement}
A basic problem in quantum information is that of distinguishing quantum states. We now describe a useful tool in this context, namely a measurement that is guaranteed to have a ``pretty good'' success probability to correctly identify an unknown state from a known ensemble.\\
Suppose that Alice (A) chooses one among $m$ pure states $|\psi_i\rangle\in\IC^d$ according to probabilities $p_i\in [0,1]$, where $p_i\geq 0$ and $\sum\limits_{i=1}^m p_i =1$, and then sends the state to Bob (B). B wants to identify the state by performing a POVM measurement $\mathcal{A}$. Let $\mathcal{E}=\lbrace(p_i,|\psi_i\rangle)\rbrace_{i=1\ldots,m}$ be the ensemble describing A's preparation procedure, denote B's optimal success probability by $P^{opt} := \max\limits_{POVM\ \mathcal{A}} P^\mathcal{A}$, where $P^\mathcal{A}:=\sum\limits_{i=1}^m p_i\langle\psi_i|A_i|\psi_i\rangle$ for a POVM $\mathcal{A}=\lbrace A_i\rbrace_{i=1,\ldots,m}$. Paul Hausladen and William Wootters \cite{Hausladen.1994} suggested a canonical form for a measurement for state discrimination, which is now usually referred to as the ``pretty good measurement'' (PGM) corresponding to the ensemble $\mathcal{E}$. It is defined in the following way:\\
First let $|\psi'_i\rangle:=\sqrt{p_i}|\psi_i\rangle$ be the states renormalized according to their respective probabilities. The density operator of the ensemble $\mathcal{E}$ is $\rho:=\sum\limits_{i=1}^m p_i|\psi_i\rangle\langle\psi_i|=\sum\limits_{i=1}^m |\psi'_i\rangle\langle\psi'_i|$. Now define $|\varphi_i\rangle:=\rho^{-\frac{1}{2}}|\psi_i\rangle$, where the inverse square root is taken only over non-zero eigenvalues of $\rho$. Now the PGM is $\mathcal{A}^{PGM}=\lbrace|\varphi_i\rangle\langle\varphi_i|\rbrace_{i=1,\ldots,m}$. (Observe that this is indeed a valid POVM, even a projection-valued measure (PVM), because $\sum\limits_{i=1}^m |\varphi_i\rangle\langle\varphi_i| = \rho^{-\frac{1}{2}} \rho \rho^{-\frac{1}{2}} = \mathds{1}_d$.)\\

The ``pretty good'' performance of the PGM was proved in \cite{Barnum.20000423}:
\begin{theorem}\label{ThmPGMSuccessProb}
For the PGM measurement defined above it holds that
\begin{align*}
P^{opt}(\mathcal{E})^2 \leq P^{PGM}(\mathcal{E}) \leq P^{opt}(\mathcal{E}).
\end{align*}
\end{theorem}

Another useful property of the PGM is that the corresponding success probability can be computed from the Gram matrix of the ensemble as follows:
\begin{lemma}\label{LmmPGMProbGram}
The success probability for the PGM measurement for an ensemble $\mathcal{E}=\lbrace(p_i,|\psi_i\rangle)\rbrace_{i=1\ldots,m}$ can be written as
\begin{align*}
	P^{PGM}(\mathcal{E})
	&= \sum\limits_{i=1}^m \sqrt{G}(i,i)^2,
\end{align*}
where $G$ is the Gram matrix with entries $G(i,j)=\sqrt{p_i p_j}\langle\psi_i|\psi_j\rangle$ for $1\leq i,j\leq m$.
\end{lemma}
\begin{proof}
This result can be shown by direct computation using the definition of the PGM and the uniqueness of the positive square root of a positive matrix.
\end{proof}

\section{The Learning Problem}\label{SctLearningProblem}
We now describe the learning task which we aim to understand. For $\ensuremath{a\in\lbrace0,1\rbrace^{n}}$, define 
\begin{align*}
f^{(a)}:\lbrace-1,1\rbrace^{n}\to\lbrace0,1\rbrace,\ f^{(a)}(x):=\sum\limits_{i=1}^{n} a_{i}\frac{1-x_{i}}{2}\pmod 2.
\end{align*}
When we observe that $\frac{1-x_{i}}{2}$ is simply the bit-description of $x_i$, it becomes clear that $f^{(a)}$ computes the parity of the entries of the bit-description of $x_i$ at the positions at which $a$ has a $1$-entry. To ease readability, we will write $\tilde{x}_i = \frac{1-x_i}{2}$.\\

The classical task which inspires our problem is the following: Given a set of
$m$ labelled examples $S=\lbrace(x_{i},f^{(a)}(x_{i}))\rbrace_{i=1}^{m}$,
where the $x_{i}$ are drawn i.i.d.~according to $D_{\mu}$, determine
the string $a$ with high success probability. Here, we assume prior
knowledge of the underlying distribution and that the underlying distribution is a $c$-bounded product distribution as introduced in subsection \ref{SbSctQuFourierSampling}. This means that we are considering a problem of exact learning from examples with instances drawn from a distribution that is known to the learner in advance.\\
Classically, as we show in section \ref{SctSampCompLowerBound}, successfully solving the task requires a number of examples that grows at least linearly in $n$. If we consider a version of this problem with noisy training data, then known classical algorithms perform worse both w.r.t.~sample complexity and running time. E.g.,~\cite{Lyubashevsky.2005} exhibits an algorithm with polynomial (superlinear) sample complexity but barely subexponential runtime (both w.r.t.~n).\\

The step to the quantum version of this problem now is the same as from classical to quantum exact learning. This means that training data is given as $m$ copies of the quantum example state $|\psi_{a}\rangle =\sum\limits_{x\in\lbrace-1,1\rbrace^{n}}\sqrt{D_{\mu}(x)}|x,f^{(a)}(x)\rangle$ and the learner is allowed to use quantum computation to process the training data. The goal of the quantum learner remains that of outputting the unknown string $a$ with high success probability.

\section{A Generalized Bernstein-Vazirani Algorithm}\label{SctGenBV}
To understand how $\mu$-biased quantum Fourier sampling can help us with this learning problem we first compute the $\mu$-biased Fourier coefficients of $g^{(a)}:=(-1)^{f^{(a)}}$, with $f^{(a)}$ for $a\in\lbrace 0,1\rbrace^n$ the linear functions defined in section $3$.

\begin{lemma}\label{LmmBiasedFourierCoeff}
Let $a\in\lbrace 0,1\rbrace^n$, $g^{(a)}:=(-1)^{f^{(a)}}$ and $\mu\in (-1,1)^n$. Then the $\mu$-biased Fourier coefficients of $g^{(a)}$ satisfy:
\begin{enumerate}[(i)]
\item If $\exists\ 1\leq i\leq n$ s.t.~$a_i=0\neq j_i$, then $\hat{g}_\mu^{(a)}(j)=0.$
\item If for all $1\leq i\leq n$ s.t.~$a_i=0$ also $j_i=0$, then
\begin{align*}
\hat{g}_\mu^{(a)}(j)=\left(\prod\limits_{l:a_l=1\neq j_l}\mu_l\right)\left(\prod\limits_{l:a_l=1= j_l}\sqrt{1-\mu_l^2}\right).
\end{align*}
\end{enumerate}
We can reformulate this as
\begin{align*}
\hat{g}_\mu^{(a)}(j) = \left(\prod\limits_{l:a_l=0} (1-j_l) \right)\left(\prod\limits_{l:a_l=1}\left((1-j_l)\mu_l + j_l \sqrt{1-\mu_l^2}\right)\right), \quad j\in\{ 0,1\}^n.
\end{align*}
\end{lemma}
\begin{proof}
We first observe that all the ``objects of interest,'' namely the probability distribution $D_\mu$, the basis functions $\phi_{\mu,j}$, and the target function $\hat{g}_\mu^{(a)}$, factorize. This now implies that also the $\mu$-biased Fourier coefficients factorize, i.e.~we have
\begin{align*}
\hat{g}_\mu^{(a_1\ldots a_n)}(j_1\ldots j_n) = \prod\limits_{i=1}^n\mathbb{E}_{D_{\mu_i}}[\phi_{\mu_i,j_i}(x_i)\cdot (-1)^{a_i\cdot \tilde{x}_i}].
\end{align*}
Therefore we only have to study the case $n=1$ in detail and the general result then follows. In this case we have $f^{(a)}(x)=a\tilde{x}$, $g^{(a)}(x)=(-1)^{a\tilde{x}}$ for $\tilde{x}=\frac{1-x}{2}$, $\phi_{\mu,0}(x)=1$, and $\phi_{\mu,1}(x)=\frac{x-\mu}{\sqrt{1-\mu^2}}$. (We leave out unnecessary indices to improve readability.) We compute
\begin{align*}
\hat{g}_\mu^{(a)}(j) 
&= \mathbb{E}_{D_\mu}[(-1)^{a\tilde{x}} \phi_{\mu,j}(x)]
=\frac{1+\mu}{2}\cdot 1\cdot \phi_{\mu,j}(1) + \frac{1-\mu }{2}\cdot (-1)^a \cdot \phi_{\mu,j}(-1).
\end{align*}
By plugging in we now obtain
\begin{align*}
\hat{g}_\mu^{(0)} (0) &= 1,\quad 
\hat{g}_\mu^{(0)} (1) = 0,\quad
\hat{g}_\mu^{(1)} (0) = \mu,\quad
\hat{g}_\mu^{(1)} (1) = \sqrt{1-\mu^2},
\end{align*}
which is exactly the claim for $n=1$.
\end{proof}

For clarity, we write down explicitly the algorithm which we obtain as a generalization of the Bernstein-Vazirani algorithm to a $\mu$-biased product distribution as Algorithm \ref{AlgBiasBV}. The generalization compared to the standard Bernstein-Vazirani algorithm consists only in going from the uniform to a more general product distribution, which gives rise to different observation probabilities.\\

\begin{algorithm}
\caption{Generalised Bernstein-Vazirani algorithm}\label{AlgBiasBV}
\begin{flushleft}
\hspace*{\algorithmicindent} \textbf{Input}: $|\psi_{a}\rangle = \sum\limits_{x\in\lbrace -1,1\rbrace^n}\sqrt{D_\mu (x)}|x,f^{(a)}(x)\rangle$ for $a\in\lbrace0,1\rbrace^{n}$, and $\mu\in [-1,1]^n$\\
\hspace*{\algorithmicindent} \textbf{Output}: $o\in\lbrace0,1\rbrace^{n}$ with probability $$\left(\prod\limits_{l:a_l=0} (1-o_l)\right)\left(\prod\limits_{l:a_l=1} \left( (1-o_l)\mu_l^2 + o_l (1-\mu_l^2)\right) \right)$$ \\
\hspace*{\algorithmicindent} \textbf{Success Probability}: $\frac{1}{2}$ 
\end{flushleft}
\begin{algorithmic}[1]
\State Perform the $\mu$-biased QFT $H_\mu$ on the first $n$ qubits, obtain the state $(H_\mu \otimes \mathds{1})|\psi_{a}\rangle$.
\State Perform a Hadamard gate on the last qubit, obtain the state $(H_\mu \otimes H)|\psi_{a}\rangle$.
\State Measure each qubit in the computational basis and observe outcome $j = j_1\ldots j_{n+1}$.
\If{$j_{n+1} = 0$}\Comment{This corresponds to a failure of the algorithm.}
	\State Output $o = \perp$. 
\ElsIf{$j_{n+1} = 1$}\Comment{This corresponds to a success of the algorithm.}
	\State Output $o = j_1\ldots j_n$.
\EndIf
\end{algorithmic}
\end{algorithm}

We now show that the output probabilities of Algorithm \ref{AlgBiasBV} are as claimed in its description. This follows directly by combining Lemma \ref{LmmBiasedQFourierSampling} on the workings of $\mu$-biased quantum Fourier sampling with Lemma \ref{LmmBiasedFourierCoeff} on the $\mu$-biased Fourier coefficients of our target functions and is the content of the following

\begin{theorem}\label{ThmBiasBVProb}
Let $|\psi_{a}\rangle =\sum\limits_{x\in\lbrace-1,1\rbrace^{n}}\sqrt{D_{\mu}(x)}|x,f^{(a)}(x)\rangle$ be a quantum example state, with $a\in\lbrace 0,1\rbrace^n$ and $\mu\in (-1,1)^n$. Then step $3$ of Algorithm \ref{AlgBiasBV} provides an outcome $|j_1\ldots j_{n+1}\rangle$ with the following properties:
\begin{enumerate}[(i)]
\item $\mathbb{P}[j_{n+1}=0]=\frac{1}{2}=\mathbb{P}[j_{n+1}=1],$
\item $\mathbb{P}[j_1\ldots j_n = a|j_{n+1}=1] = \prod\limits_{l:a_l=1} (1-\mu_l^2)$,
\item for $o\neq a$: $$\mathbb{P}[j_1\ldots j_n = o|j_{n+1}=1] = \prod\limits_{l:a_l=0} (1-o_l)\cdot\prod\limits_{l:a_l=1} \big( (1-o_l)\mu_l^2 + o_l (1-\mu_l^2)\big),$$
\item $\mathbb{P}[\exists 1\leq i\leq n: a_i = 0 \neq j_i | j_{n+1} = 1] =0,$ and
\item $\mathbb{P}[\exists 1\leq i\leq n: a_i = 1 \neq j_i | j_{n+1} = 1] \leq\sum\limits_{i=1}^n \mu_i^2$. In particular, if $D_\mu$ is $c$-bounded, then $\mathbb{P}[\exists 1\leq i\leq n: a_i = 1 \neq j_i | j_{n+1} = 1] \leq n(1-c)^2$.
\end{enumerate}
\end{theorem}

Note that $(v)$ can be trivial if the bias is too strong. This observation already hints at why we later use different procedures for arbitrary and for small bias.\\
We also want to point out that in the case of no bias (i.e.~$\mu=0$), Algorithm \ref{AlgBiasBV} simply reduces to the well-known Berstein-Vazirani algorithm \cite{Bernstein.1993}.

\section{Quantum Sample Complexity Upper Bounds}\label{SctQuantSampCompUpperBounds}
This section contains the description of two procedures for solving the task of learning an unknown Boolean linear function from quantum examples w.r.t.~a product distribution. (Here, we assume perfect quantum examples, noisy examples will be taken into consideration in the next section.) It is subdivided into an approach which is applicable for arbitrary (albeit not full) bias in the product distribution and a strategy which produces better results but is only valid for small bias.

\subsection{Arbitrary Bias}
As in the case of learning w.r.t.~the uniform distribution we intend to run the generalized Bernstein-Vazirani algorithm multiple times as a subroutine and then use our knowledge of the outcome of the subroutine together with probability-theoretic arguments. The main difficulty compared to the case of an example state arising from the uniform distribution lies in the fact that, whereas an observation of $j_{n+1}=1$ when performing the standard Bernstein-Vazirani algorithm guarantees that $j_1\ldots j_n$ equals the desired string, this is not true in the $\mu$-biased case. Hence, we have to develop a different procedure of learning from the outcomes of the subroutine. For this purpose we propose Algorithm \ref{AlgAmpBiasBV1}.\\

\begin{algorithm}
\caption{Amplified Generalised Bernstein-Vazirani algorithm - Version $1$}\label{AlgAmpBiasBV1}
\begin{flushleft}
\hspace*{\algorithmicindent} \textbf{Input}: $m$ copies of $|\psi_{a}\rangle = \sum\limits_{x\in\lbrace -1,1\rbrace^n}\sqrt{D_\mu (x)}|x,f^{(a)}(x)\rangle$ for $a\in\lbrace0,1\rbrace^{n}$, where the number of copies is  $m\geq C \left(\left\lceil \left(2\ln\left(\frac{1}{1-c+\frac{c^2}{2}}\right)\right)^{-1} \left(\ln (n)+ \ln(\frac{2}{\delta})\right)\right\rceil\right)$ for a suitable constant $C>0$, and $\mu\in (-1,1)^n$ and $c\in (0,1]$ s.t.~$D_\mu$ is $c$-bounded.\\
\hspace*{\algorithmicindent} \textbf{Output}: $a\in\lbrace0,1\rbrace^{n}$\\
\hspace*{\algorithmicindent} \textbf{Success Probability}: $\geq 1-\delta$
\end{flushleft}
\begin{algorithmic}[1]
\For{$1\leq l \leq m$}
	\State Run Algorithm \ref{AlgBiasBV} on the $l^{th}$ copy of $|\psi_{a}\rangle$, store the output as $o^{(l)}$.
\EndFor
\If{$\exists 1\leq l\leq m: o^{(l)}\neq\perp$}
	\For{$1\leq i\leq n$}
		\State Let $o_i := \max\limits_{l:o^{(l)}\neq\perp} o_i^{(l)}$.
	\EndFor
	\State Output $o = o_1\ldots o_n$.
\ElsIf{$\forall 1\leq l\leq m: o^{(l)}=\perp$}
	\State Output $o = \perp$.
\EndIf
\end{algorithmic}
\end{algorithm}

The amplification procedure in Algorithm \ref{AlgAmpBiasBV1} differs from the majority vote in the standard Bernstein-Vazirani learning procedure (w.r.t.~the uniform distribution) as used in \cite{Cross.2015} and \cite{Grilo.20180410} in the following two ways: Instead of working on the level of the whole string, we use a componentwise strategy. And instead of taking a majority vote over observed values, we take a maximum to account for the asymmetry in the probability of an observation error (see Theorem \ref{ThmBiasBVProb}). \\

We now show that the number of copies postulated in Algorithm \ref{AlgAmpBiasBV1} is actually sufficient to achieve the desired success probability.
\newpage
\begin{theorem}\label{ThmAmpBiasBV1}
Let $|\psi_{a}\rangle =\sum\limits_{x\in\lbrace-1,1\rbrace^{n}}\sqrt{D_{\mu}(x)}|x,f^{(a)}(x)\rangle$, $a\in\lbrace 0,1\rbrace^n$, $\mu\in (-1,1)^n$ s.t.~$D_\mu$ is $c$-bounded for some $c\in (0,1]$. Then $$\mathcal{O}\left(\left(2\ln\left(\frac{1}{1-c+\frac{c^2}{2}}\right)\right)^{-1} \left(\ln (n)+ \ln(\frac{2}{\delta})\right)\right)$$ copies of the quantum example state $|\psi_{a}\rangle$ are sufficient to guarantee that, with probability $\geq 1-\delta$, Algorithm \ref{AlgAmpBiasBV1} outputs the string $a$.
\end{theorem}

\begin{proof}
We want to show that $\mathbb{P}[\textrm{Algorithm \ref{AlgAmpBiasBV1} does not output } a] \leq \delta$. We do so by treating separately the cases in which Algorithm \ref{AlgAmpBiasBV1} does not output $a$.\\
The first such case occurs if $o=\perp$. The second such case would be that there exists $1\leq i\leq n$ s.t.~$a_i = 0 \neq o_i$, but due to Theorem \ref{ThmBiasBVProb}, this is an event of probability $0$. The third and last such case is that there exists $1\leq i\leq n$ s.t.~$a_i = 1 \neq o_i$. Hence, we can decompose the probability of Algorithm \ref{AlgAmpBiasBV1} producing a wrong output as
\begin{align}
&\mathbb{P}[\textrm{Algorithm \ref{AlgAmpBiasBV1} does not output } a] \nonumber\\
&=\mathbb{P}[\textrm{Algorithm \ref{AlgAmpBiasBV1} outputs }\perp] + \mathbb{P}[\exists 1\leq i\leq n: a_i = 1 \neq o_i]. \label{eq:ProbSum}
\end{align}
First, we bound the probability of the algorithm outputting $\perp$ (i.e.~of each subroutine failing) as follows:
\begin{align*}
&\mathbb{P}[\textrm{Algorithm \ref{AlgAmpBiasBV1} outputs } \perp]\\
=~ &\mathbb{P}[\forall 1\leq l\leq m: \textrm{Algorithm \ref{AlgBiasBV} applied to }|\psi_{a}\rangle\textrm{ outputs }\perp]\\
=~ &\left(\frac{1}{2}\right)^m,
\end{align*}
where the last step uses Theorem \ref{ThmBiasBVProb} and that the training data consists of independent copies of $|\psi_a\rangle$, i.e.~is given as a product state. The choice of $m$ now guarantees that this last term is $\leq\frac{\delta}{2}$ (if we choose the constant $C>0$ sufficiently large).\\

Now we bound the second term in equation (\ref{eq:ProbSum}). We make the following observation: Suppose $1\leq i\leq n$ is s.t.~$a_i=1$. As the Fourier coefficients, and with them the output probabilities, factorize, the probability of Algorithm \ref{AlgBiasBV} outputting a string $j_1\ldots j_n$ with $j_i = 1 = a_i$ is simply the probability of Algorithm \ref{AlgBiasBV} applied to only the subsystem state of $|\psi_{a}\rangle$ corresponding to the $i^{th}$ and the $(n+1)^{st}$ subsystem outputting a $1$. By Theorem \ref{ThmBiasBVProb}, this probability is 
\begin{align*}
\mathbb{P}[j_i = 1]
= \mathbb{P}[j_{n+1} = 1]\cdot\mathbb{P}[j_i = 1|j_{n+1}=1] 
= \frac{1}{2}\cdot (1-\mu_i^2).
\end{align*}
Hence, assuming $a_i=1$, the probability of not observing a $1$ at the $i^{th}$ position in any of the $m$ runs of Algorithm \ref{AlgBiasBV} is $\left( 1-\frac{1}{2}\cdot (1-\mu_i^2)\right)^m = \left(\frac{1}{2}(1+\mu_i^2)\right)^m$. By $c$-boundedness of the distribution $D_\mu$ we get
\begin{align*}
\left(\frac{1}{2}(1+\mu_i^2)\right)^m
\leq \left(\frac{1}{2} + \frac{1}{2}(1-c)^2\right)^m
= \left (1- c +\frac{c^2}{2}\right)^m.
\end{align*}
So using the union bound we arrive at
\begin{align*}
&\mathbb{P}[\exists 1\leq i\leq n: a_i = 1 \neq o_i]\\
&= \mathbb{P}[\exists 1\leq i\leq n: a_i = 1 \textrm{ and in }m\textrm{ runs no }1\textrm{ is observed at the }i^{th}\textrm{ entry}]\\
&\leq \sum\limits_{i=1}^n\mathbb{P}[a_i = 1 \textrm{ and in }m\textrm{ runs no }1\textrm{ is observed at the }i^{th}\textrm{ entry}]\\
&\leq n\cdot\left (1- c +\frac{c^2}{2}\right)^m.
\end{align*}
The choice of $m$ guarantees that this last term is $\leq\frac{\delta}{2}$ (if we choose the constant $C>0$ sufficiently large).\\

We now combine this with equation (\ref{eq:ProbSum}) and obtain
\begin{align*}
\mathbb{P}[\textrm{Algorithm \ref{AlgAmpBiasBV1} does not output } a] 
&\leq \frac{\delta}{2} + \frac{\delta}{2} = \delta,
\end{align*}
which finishes the proof.
\end{proof}

\begin{remark}
We want to comment shortly on the dependence of the sample complexity bound on the $c$-boundedness constant by considering extreme cases. As $c\to 0$, i.e.~we allow more and more strongly biased distributions, the sample complexity goes to infinity. This reflects the fact that in the case of a fully biased underlying product distribution, only a single bit of information about $a$ can be extracted, so exactly learning the string $a$ is (in general) not possible.\\
For $c=1$, i.e.~the case of no bias, we simply obtain that $\mathcal{O}\left(\left(\ln (n)+ \ln(\frac{2}{\delta})\right)\right)$ copies of the quantum example state are sufficient. Note that this does not coincide with the bound obtained for the standard Bernstein-Vazirani procedure which is independent of $n$. (This can easily be shown using Lemma \ref{LmmHoeffding}.)\\
This discrepancy is due to the difference in ``amplification procedures.'' Namely, in Algorithm \ref{AlgAmpBiasBV1} we do not explicitly make use of the knowledge that, given $j_{n+1}=1$, we know the probability of $j_1\ldots j_n=a_1\ldots a_n$ because, whereas for $\mu=0$ this probability equals $1$, for $\mu\neq 0$ it can become small. Hence, for $\mu\neq 0$ our algorithm introduces an additional procedure to deal with the uncertainty of $j_1\ldots j_n$ even knowing $j_{n+1}$ and we see in the proof that this yields the additional $\ln(n)$ term. In the next subsection we describe a way to get rid of exactly that $\ln(n)$ term for ``small'' bias.
\end{remark}

\subsection{Small Bias}
In this subsection we want to study the case in which $(v)$ of Theorem \ref{ThmAmpBiasBV1} gives a good bound. Namely, throughout this subsection we will assume that the $c$-boundedness constant is s.t.~$n(1-c)^2<\frac{1}{2}$ or, equivalently, $c > 1 - \frac{1}{\sqrt{2n}}$. This assumption will allow us to apply a different procedure to learn from the output of Algorithm \ref{AlgBiasBV} and thus obtain a different bound on the sample complexity of the problem. Note, however, that this requirement becomes more restrictive with growing $n$ and can in the limit $n\to\infty$ only be satisfied by $c=1$, i.e.~for the underlying distributions being uniform. Also, we will from now on refer to $c$ as $c$-boundedness parameter because the name ``constant'' would hide the $n$-dependence.\\

Our procedure for the case of small bias is given in Algorithm \ref{AlgAmpBiasBV2}.

\begin{algorithm}
\caption{Amplified Generalised Bernstein-Vazirani algorithm - Version $2$}\label{AlgAmpBiasBV2}
\begin{flushleft}
\hspace*{\algorithmicindent} \textbf{Input}: $m$ copies of $|\psi_{a}\rangle = \sum\limits_{x\in\lbrace -1,1\rbrace^n}\sqrt{D_\mu (x)}|x,f^{(a)}(x)\rangle$ for $a\in\lbrace0,1\rbrace^{n}$, where the number of copies is  $m\geq C\left(\frac{4}{(1-2n(1-c)^2)^2}\ln\left(\frac{2}{\delta}\right)\right)$, as well as $\mu\in [-1,1]^n$ and $c\in (0,1]$ s.t.~$D_\mu$ is $c$-bounded.\\
\hspace*{\algorithmicindent} \textbf{Output}: $a\in\lbrace0,1\rbrace^{n}$\\
\hspace*{\algorithmicindent} \textbf{Success Probability}: $\geq 1-\delta$
\end{flushleft}
\begin{algorithmic}[1]
\For{$1\leq l \leq m$}
	\State Run Algorithm \ref{AlgBiasBV} on the $l^{th}$ copy of $|\psi_{a}\rangle$, store the output as $o^{(l)}$.
\EndFor
\If{$\exists 1\leq l\leq m: o^{(l)}\neq\perp$}
	\For{$1\leq i\leq n$}
		\State Let $o_i = \argmax\limits_{r\in\lbrace 0,1\rbrace} \lvert\lbrace 1\leq l\leq m | o^{(l)}_i = r\rbrace\rvert$.
	\EndFor
	\State Output $o=o_1 \ldots o_n$.
\ElsIf{$\forall 1\leq l\leq m: o^{(l)}=\perp$}
	\State Output $o=\perp$.
\EndIf
\end{algorithmic}
\end{algorithm}

\begin{theorem}\label{ThmAmpBiasBV2}
Let $|\psi_{a}\rangle =\sum\limits_{x\in\lbrace-1,1\rbrace^{n}}\sqrt{D_{\mu}(x)}|x,f^{(a)}(x)\rangle$, $a\in\lbrace 0,1\rbrace^n$, $\mu\in (-1,1)^n$ s.t.~$D_\mu$ is $c$-bounded for some $c\in (0,1]$ satisfying $c > 1 - \frac{1}{\sqrt{2n}}$. Then $$\mathcal{O}\left(\frac{1}{(1-2n(1-c)^2)^2} \ln\left(\frac{1}{\delta}\right)\right)$$ copies of the quantum example state $|\psi_{a}\rangle$ are sufficient to guarantee that, with probability $\geq 1-\delta$, Algorithm \ref{AlgAmpBiasBV2} outputs the string $a$.
\end{theorem}

Note that due to the required lower bound on $c$ the sample complexity upper bound basically loses its $n$-dependence. This is different from the result of Theorem \ref{ThmAmpBiasBV1}, where $n$ explicitly entered the upper bound.
\\

\begin{proof}
By Theorem \ref{ThmBiasBVProb}, we have $\mathbb{P}[j_{n+1}=1]=\frac{1}{2}$. Hence, the probability of observing $j_{n+1}=1$ in at most $k-1$ of the $m$ runs of Algorithm \ref{AlgBiasBV} is given by $$\sum\limits_{l=0}^{k-1} \binom{m}{i} \left(\frac{1}{2}\right)^i \left(\frac{1}{2}\right)^{m-i} = \mathbb{P}\left[\textrm{Bin}(m,\frac{1}{2})\geq m-k\right],$$ where $\textrm{Bin}$ denotes a binomial distribution.\\

Next we assume $k\leq\frac{m}{2}$ (this will be justified later in the proof) and use Hoeffding's inequality (Lemma \ref{LmmHoeffding}) to obtain
\begin{align}
\mathbb{P}\left[\textrm{Bin}(m,\frac{1}{2})\geq m-k\right]
&= \mathbb{P}\left[\textrm{Bin}(m,\frac{1}{2}) - \frac{m}{2}\geq m - k - \frac{m}{2}\right]\nonumber\\
&\leq \exp\left(-\frac{2\left(\frac{m}{2} - k\right)^2}{m}\right). \label{eq:NoOfRuns}
\end{align}

We will now search for the number of observations of $j_{n+1}=1$ which is required to guarantee that the majority string is correct with high probability. Assume that we observe $j_{n+1}=1$ in $k$ runs of Algorithm \ref{AlgBiasBV}, $k\in 2\IN$. (The latter assumption clearly does not significantly change the number of copies.). Using $(v)$ from Theorem \ref{ThmBiasBVProb} we see that
\begin{align*}
\mathbb{P}[\exists 1\leq i\leq n: a_i \neq o_i]
&\leq \mathbb{P}[\exists 1\leq i\leq n: a_i = 0 \neq o_i]\\ 
&\hphantom{\leq} + \mathbb{P}[\exists 1\leq i\leq n: a_i = 1 \neq o_i]\\
&\leq 0 + \sum\limits_{l=\lceil\frac{k}{2}\rceil}^{k} \binom{k}{l} \cdot(1- n(1-c)^2 )^{k-l} \cdot(n(1-c)^2)^{l}\\
&= \mathbb{P}\left[\textrm{Bin}(k, n(1-c)^2) \geq \frac{k}{2}\right],
\end{align*}
where the second inequality uses that the majority string can only be wrong if in at least half of the runs where we observed $j_{n+1}=1$ there was some error in the remaining string.\\
Next we use Hoeffding's inequality and obtain, using our assumption $n(1-c)^2<\frac{1}{2}$, that
\begin{align*}
& \mathbb{P}\left[\textrm{Bin}(k, n(1-c)^2) \geq \frac{k}{2}\right] \\
&= \mathbb{P}\left[\textrm{Bin}(k, n(1-c)^2) - kn(1-c)^2 \geq \frac{k}{2} - kn(1-c)^2 \right]\\
&\leq \exp\left(-k\frac{(1-2n(1-c)^2)^2}{2}\right).
\end{align*}
We now set this last expression $\leq \frac{\delta}{2}$ for $\delta\in (0,1)$ and rearrange the inequality to
\begin{align}
k\geq \frac{2}{(1-2n(1-c)^2)^2}\ln\left(\frac{2}{\delta}\right). \label{eq:NoOf1s}
\end{align}
Combining equations (\ref{eq:NoOf1s}) and (\ref{eq:NoOfRuns}) we now require 
\begin{align*}
\exp\left(-\frac{2\left(\frac{m}{2} - \frac{2}{(1-2n(1-c)^2)^2}\ln\left(\frac{2}{\delta}\right)\right)^2}{m}\right) \overset{!}{\leq} \frac{\delta}{2}.
\end{align*}
Rearranging this inequality gives
\begin{align*}
m^2 - 2m\left(\left(\tfrac{1-2n(1-c)^2}{2}\right)^{-2} - 1\right) \ln\left(\tfrac{2}{\delta}\right) + \left(\tfrac{1-2n(1-c)^2}{2}\right)^{-4}\ln^2\left(\tfrac{2}{\delta}\right)\geq 0.
\end{align*}
By finding the zeros of this quadratic function we get to the sufficient sample size
\begin{align*}
m \geq &\left(\left(\tfrac{1-2n(1-c)^2}{2}\right)^{-2} - 1\right)\ln\left(\tfrac{2}{\delta}\right)  \\
&+\sqrt{\left(\left(\left(\tfrac{1-2n(1-c)^2}{2}\right)^{-2} - 1\right)\ln\left(\tfrac{2}{\delta}\right)\right)^2
- \left(\tfrac{1-2n(1-c)^2}{2}\right)^{-4}\ln^2\left(\tfrac{2}{\delta}\right)}.
\end{align*}
This is in particular guaranteed if
\begin{align*}
m \geq \frac{4}{(1-2n(1-c)^2)^2} \ln\left(\frac{2}{\delta}\right).
\end{align*}
Note that this lower bound in particular implies $m\geq 2k$, as required earlier in the proof. This proves the claim of the theorem thanks to the union bound.
\end{proof}

Morally speaking, Theorem \ref{ThmAmpBiasBV2} shows that for product distributions which are close enough to the uniform distribution the sample complexity upper bound is the same as for the unbiased case. We conjecture that there is an explicit noise threshold above which this sample complexity cannot be reached (see the discussion in section $\ref{SctSampCompLowerBound}$), but have not yet succeeded in identifying such a critical value.\\

In this section, we have discussed the case of quantum training data that perfectly represents the target function in a superposition state. Similar results can be proved in the case of noisy quantum training data. As the reasoning is analogous to the one presented here, the details are deferred to Appendix \ref{SctNoiseStability}.

\section{Sample Complexity Lower Bounds}\label{SctSampCompLowerBound}
After proving upper bounds on the number of required quantum examples by exhibiting explicit learning procedures in the previous section, we now study the converse question of sample complexity lower bounds. We will prove both classical and quantum sample complexity lower bounds and then relate them to the above results. Our proof strategy follows a state-discrimination-based strategy from \cite{Arunachalam.2018}.

\subsection{Classical Sample Complexity Lower Bounds}
We first prove a sample complexity lower bound for the classical version of our learning problem that upon comparison with our obtained quantum sample complexity upper bounds shows the advantage of quantum examples over classical training data in this setting. Neither the result nor the proof strategy are new but we include them for completeness.

\begin{theorem}\label{ThmSampCompLowerBoundClass}
Let $a\in\lbrace 0,1\rbrace^n$, $\mu\in (-1,1)^n$ s.t.~$\mu$ is $c$-bounded for some $c\in (0,1]$. Let $\mathcal{A}$ be a classical learning algorithm and let $m\in\IN$ be such that upon input of $m$ examples of the form $(x_i,f^{(a)}(x_i))$, with $x_i$ drawn i.i.d.~according to $D_\mu$, with probability $\geq 1-\delta$ w.r.t.~the choice of training data, $\mathcal{A}$ outputs the string $a$. Then $m\geq\Omega (n)$.
\end{theorem}

\begin{proof}
Let $A$ be a random variable uniformly distributed on $\lbrace 0,1\rbrace^n$. ($A$ describes the underlying string from the initial perspective of the learner.) Let $B=(B_1,\ldots, B_m)$ be a random variable describing the training data corresponding to the underlying string. Our proof will have three main steps: First, we prove a lower bound on $I(A:B)$ from the learning requirement. Second, we observe that $I(A:B)\leq m\cdot I(A:B_1)$. And third, we prove an upper bound on $I(A:B_1)$. Then combining the three steps will lead to a lower bound on $m$.\\
We start with the mutual information lower bound. Let $h(B)\in\lbrace 0,1\rbrace^n$ denote the random variable describing the output hypothesis of the algorithm $\mathcal{A}$ upon input of training data $B$. Let $Z=\mathds{1}_{ \{h(B)=A\} }$. By the learning requirement we have $\mathbb{P}[Z=1]\geq 1-\delta$ and thus $H(Z)\leq H(\delta)$. Therefore we obtain
\begin{align*}
I(A:B) &= H(A) - H(A|B)\\
&\geq H(A) - H(A|B,Z) - H(Z)\\
&= H(A) - \mathbb{P}[Z=1] H(A|B,Z=1) - \mathbb{P}[Z=0] H(A|B,Z=0) - H(Z)\\
&\geq n - \mathbb{P}[Z=1]\cdot 0 - \delta n - H(\delta)\\
&= (1-\delta)n - H(\delta)\\
&= \Omega(n).
\end{align*}
We now show that from $m$ examples we can gather at most $m$ times as much information as from a single example. Here we directly cite from \cite{Arunachalam.2018}. Namely,
\begin{align*}
I(A:B)&= H(B) - H(B|A)
= H(B) - \sum\limits_{i=1}^m H(B_i|A)\\
&\leq \sum\limits_{i=1}^m H(B_i) - H(B_i|A)
= \sum\limits_{i=1}^m I(A:B_i)
= m\cdot I(A:B_1).
\end{align*}
Here, the second step uses independence of the $B_i$ conditioned on $A$, the third step uses subadditivity of the Shannon entropy and the final step uses that the distributions of $(A,B_i)$ are the same for all $1\leq i\leq m$.\\
We come to the upper bound on the mutual information. Write $B_1=(X,L)$ for $X\in\lbrace -1,1\rbrace^n$ and $L\in\lbrace 0,1\rbrace$, i.e. with probability $D_\mu (x)$ we have $(X,L)=(x,f^{(a)}(x))$. Note that $I(A:X)=0$ because $X$ and $A$ are independent random variables. Also, $I(A:L|X=1\ldots 1)=0$ because $f^{(a)}(1\ldots 1)=0\ \forall a\in\lbrace 0,1\rbrace^n$, and for $x\in\lbrace -1,1\rbrace^n\setminus\lbrace 1\ldots 1\rbrace$
\begin{align*}
I(A:L|X=x) &= I(A_{\lbrace i| X_i=-1\rbrace}:L|X=x)\\
&= H(A_{\lbrace i|X_i=-1\rbrace}|X=x) - H(A_{\lbrace i|X_i=-1\rbrace}|L,X=x)\\
&= \left\lvert\lbrace i|x_i=-1\rbrace\right\rvert - (\left\lvert\lbrace i|x_i=-1\rbrace\right\rvert - 1)\\
&= 1.
\end{align*}
Here, the first step is due to the fact that $f^{(a)}(x)$ does not depend on the entries $a_j$ with $x_j=1$, the third step follows because $A_{\lbrace i|x_i=-1\rbrace}$ is uniformly distributed on a set of size $2^{\left\lvert\lbrace i|x_i=-1\rbrace\right\rvert}$ and $f^{(a)}$ assigns the labels $0$ and $1$ to half of the elements of that set, respectively.\\
This now implies
\begin{align*}
I(A:B_1) &= I(A:X) + I(A:L|X)\\
&= 0 + \sum\limits_{x\in\lbrace -1,1\rbrace^n} D_\mu (x) I(A:L|X=x)\\
&= 1.
\end{align*}
Here, the first step is due to the chain rule for mutual information and the last step simply uses the fact that $D_\mu$ defines a probability distribution.\\
Now we combine our upper and lower bounds on the mutual information and obtain
\begin{align*}
m\geq (1-\delta)n - H(\delta) = \Omega(n),
\end{align*}
as claimed.
\end{proof}

\begin{remark}
The result of Theorem \ref{ThmSampCompLowerBoundClass} is intuitively clear: In order to identify the underlying string the learning algorithm has to learn $n$ bits of information. However, a condition of the form $f^{(a)}(x)=l$ for $x\in\lbrace 0,1\rbrace^n, l\in\lbrace 0,1\rbrace$, takes away at most one degree of freedom from the initial space $\lbrace 0,1\rbrace^n$ for $a$ and thus from such an equality the algorithm can extract at most $1$ bit of information. So at least $n$ examples will be required. This observation is thus neither new nor surprising. But we want to emphasize that this analysis works independently of the product structure of the underlying distribution $D_\mu$.
\end{remark}

If we compare the classical lower bound from Theorem \ref{ThmSampCompLowerBoundClass} with our quantum upper bounds from Theorems \ref{ThmAmpBiasBV1} and \ref{ThmAmpBiasBV2}, we conclude that quantum examples allow us to strictly outperform the best possible classical algorithm w.r.t.~the number of required examples.

\subsection{Quantum Sample Complexity Lower Bounds}
We can use a similar argument to prove quantum sample complexity lower bounds. Note that steps $1$ and $2$ carry over with (almost) no changes. Only the analysis of step $3$ changes significantly. Even though this proof strategy is possible, as in \cite{Arunachalam.2018} it can be improved upon by an argument based on state discrimination. We will thus follow this same approach.\\

An $n$-independent quantum sample complexity lower bound is given in the following 

\begin{lemma}\label{LmmQuSampLowerBoundDelta}
Let $|\psi_{a}\rangle =\sum\limits_{x\in\lbrace-1,1\rbrace^{n}}\sqrt{D_{\mu}(x)}|x,f^{(a)}(x)\rangle$, $a\in\lbrace 0,1\rbrace^n$, $\mu\in (-1,1)^n$ s.t.~$D_\mu$ is $c$-bounded for some $c\in (0,1]$. Let $\mathcal{A}$ be a quantum learning algorithm and let $m\in\IN$ be such that upon input of $m$ copies of $|\psi_{a}\rangle$, with probability $\geq 1-\delta$, $\mathcal{A}$ outputs the string $a$. Then $m\geq\Omega(\frac{1}{c}\ln(\frac{1}{\delta}))$.
\end{lemma}

\begin{remark}
Note that any quantum sample complexity lower bound will also lower bound the classical sample complexity. Hence, Lemma \ref{LmmBiasedQFourierSampling} also holds in the scenario of the previous subsection, which is why we did not discuss the $\delta$-dependence there.
\end{remark}

\begin{proof}
Let $a,b\in\lbrace 0,1\rbrace^n$ s.t.~there is exactly one $1\leq i\leq n$ s.t.~$a_i\neq b_i$. As $\mathcal{A}$ is able to distinguish the quantum states $|\psi_{a}\rangle^{\otimes m}$ and $|\psi_{b}\rangle^{\otimes m}$ with success probability $\geq 1-\delta$, we have $|\langle \psi_{a}|\psi_{b}\rangle^m|\leq 2\sqrt{\delta(1-\delta)}$ (see subsection $3.2$). We compute
\begin{align*}
\langle \psi_{a}|\psi_{b}\rangle
&= \sum\limits_{x,y\in\lbrace -1,1\rbrace^n} \sqrt{D_\mu (x) D_\mu (y)}\langle x,f^{(a)}(x)|y,f^{(b)}(y)\rangle\\
&= \sum\limits_{x\in\lbrace -1,1\rbrace^n} D_\mu (x) \delta_{f^{(a)}(x),f^{(b)}(x)}.
\end{align*}
By our assumption on $a$ and $b$, $\delta_{f^{(a)}(x),f^{(b)}(x)}\geq\delta_{x_i,1}$. Therefore
\begin{align*}
\langle \psi_{a}|\psi_{b}\rangle \geq\mathbb{P}_{D_\mu}[x_i=1] = \frac{1+\mu_i}{2}.
\end{align*}
We now combine this with our upper bound and rearrange to obtain
\begin{align*}
m&\geq \left(\ln\left(\frac{1+\mu_i}{2}\right)\right)^{-1}\left(\ln(2) + \frac{1}{2}\ln(\delta(1-\delta))\right)\\
&\geq\Omega\left(\frac{1}{\mu_i -1}\ln(\delta)\right)\\
&\geq \Omega\left(\frac{1}{c}\ln\left(\frac{1}{\delta}\right)\right),
\end{align*}
where we used the elementary inequality $\frac{1}{x-1}-\left(\ln\left(\frac{1+x}{2}\right)\right)^{-1}\geq 0$ for $x\in [0,1)$ combined with $\ln(\delta)\leq 0$.
\end{proof}

We will compare this lower bound with our upper bound(s) from section $\ref{SctQuantSampCompUpperBounds}$ later on. Now we turn to the $n$-dependent part of the sample complexity lower bound.

\begin{theorem}\label{ThmQuSampCompLowerBoundV2}
Let $|\psi_{a}\rangle =\sum\limits_{x\in\lbrace-1,1\rbrace^{n}}\sqrt{D_{\mu}(x)}|x,f^{(a)}(x)\rangle$, $a\in\lbrace 0,1\rbrace^n$, and $\mu\in (-1,1)$ be such that $\mu_i=\mu \geq 1-\frac{1}{\ln(n)}$ for all $1\leq i\leq n$. Let $\mathcal{A}$ be a quantum learning algorithm and let $m\in\IN$ be such that upon input of $m$ copies $|\psi_{a}\rangle$, with probability $\geq 1-\delta$, $\mathcal{A}$ outputs the string $a$, for $0<\delta\leq\frac{1}{3}$. Then $m\geq \Omega\left(\ln (n)\right)$.
\end{theorem}

Before going into the detailed proof, we give an overview over its underlying idea. The learning assumption implies that $\mathcal{A}$ is able to identify a state from the ensemble $\mathcal{E}=\left\lbrace \left( \frac{1}{2^n},|\psi_{a}\rangle^{\otimes m}\right)\right\rbrace_{a\in\{ 0,1\}^n}$ with success probability $\geq 1-\delta$. Thus we will obtain a lower bound on $m$ by proving an upper bound on the optimal success probability for this state identification task.\\
Recall that by Theorem \ref{ThmPGMSuccessProb}, the optimal success probability can be upper bounded by the square root of the PGM success probability. Moreover, by Lemma \ref{LmmPGMProbGram}, the latter can be computed via the Gram matrix of the ensemble. Thus, we now first study the Gram matrix and its square root and then use these results to bound the optimal success probability.\\

We first recall a well-known result on the diagonalization of matrices with a specific structure, namely matrices whose entries can be written as Boolean function of the sum of the indices.
\begin{lemma}\label{LmmDiagGramMatrix}
Let $G\in\IR^{2^n\times 2^n}$ be a matrix with entries given by $G(a,b)=g(a+b)$ for $a,b\in\{ 0,1\}^n$ and a function $g:\{ 0,1\}^n\to\IR$. Then 
\begin{align*}
	(H G H^{-1})(a,b) = 2^n \hat{g}(a)\delta_{a,b},
\end{align*}
with $H\in\IR^{2^n\times 2^n}$ given by $H(a,b)=\frac{(-1)^{a\cdot b}}{\sqrt{2^n}}$. In other words, the set of eigenvalues of $G$ is given by $\{2^n\hat{g}(a)~|~ a\in\{ 0,1\}^n \}$ and $G$ is unitarily diagonalized by $H$.
\end{lemma}
\begin{proof}
The proof can be found in \cite{Arunachalam.2018}, we reproduce it in Appendix \ref{SctProofs}.
\end{proof}

We will later apply this result for $G$ being the Gram matrix corresponding to the ensemble in our state identification task. Motivated by Lemma \ref{LmmPGMProbGram}, we first use the diagonalization of such a matrix to explicitly compute the diagonal entries of the matrix square root.

\begin{corollary}\label{CrlDiagEntriesRootGram}
Let $G\in\IR^{2^n\times 2^n}$ be a matrix with entries given by $G(a,b)=g(a+b)$ for $a,b\in\{ 0,1\}^n$ and a function $g:\{ 0,1\}^n\to\IR$. Then, for every $a\in\{ 0,1\}^n$
\begin{align*}
	\sqrt{G}(a,a) = \frac{1}{\sqrt{2^n}}\sum\limits_{j\in\{ 0,1\}^n} \sqrt{\hat{g}(j)}.
\end{align*}
\end{corollary}
\begin{proof}
The proof can be found in \cite{Arunachalam.2018}, we reproduce it in Appendix \ref{SctProofs}.
\end{proof}

With this we can now prove Theorem \ref{ThmQuSampCompLowerBoundV2}:\\
\begin{proof} (Proof of Theorem \ref{ThmQuSampCompLowerBoundV2})\\
As discussed above, we consider the problem of state identification with the ensemble $\mathcal{E}=\left\lbrace \left( \frac{1}{2^n},|\psi_{a}\rangle^{\otimes m}\right)\right\rbrace_{a\in\{ 0,1\}^n}$. By Lemma \ref{LmmPGMProbGram}, with the Gram matrix $G_m(a,b):=\frac{1}{2^n}\langle \psi_{a}|\psi_{b}\rangle^m$ we can write the success probability as 
\begin{align*}
	P^{PGM}(\mathcal{E})
	&= \sum\limits_{a\in\{ 0,1\}^n} \sqrt{G_m}(a,a)^2.
\end{align*}
In our scenario, the Gram matrix has entries
\begin{align*}
	G_m(a,b)
	&=\frac{1}{2^n}\langle \psi_{a}|\psi_{b}\rangle^m\\
	&= \frac{1}{2^{n+m}}\left( 1 + \mu^{\textrm{d}_H(a,b)}\right)^m
	= \frac{1}{2^{n+m}}\left( 1 + \mu^{\textrm{d}_H(a+b,0)}\right)^m.
\end{align*}
This can e.g.~be shown by induction on $n$ when observing that
\begin{align*}
&\mathbb{P}_{D_\mu}[f^{(a)}(x)=f^{(b)}(x)]\\
=~ &\mathbb{P}_{D_\mu}[f^{(a_{1:n-1})}(x_{1:n-1})=f^{(b_{1:n-1})}(x_{1:n-1})~\wedge ~ a_n\frac{1-x_n}{2}=b_n\frac{1-x_n}{2}]\\
 +~ &\mathbb{P}_{D_\mu}[f^{(a_{1:n-1})}(x_{1:n-1})\neq f^{(b_{1:n-1})}(x_{1:n-1})~\wedge ~ a_n\frac{1-x_n}{2}\neq b_n\frac{1-x_n}{2}].
\end{align*}
In particular, we can write $G_m(a,b)=f_m(a+b)$ for the function $f_m(x)=\frac{1}{2^{n+m}}\left( 1 + \mu^{\textrm{d}_H(x,0)}\right)^m$. From now on, we will write $|x|:=\textrm{d}_H(x,0)$. By Corollary \ref{CrlDiagEntriesRootGram}, we can upper bound the diagonal entries of $\sqrt{G_m}$ (and thus the PGM and the optimal success probability) by upper bounding the (unbiased) Fourier coefficients of $f_m$. To this end, consider for $j\in\{ 0,1\}^n$
\begin{align*}
	0\leq \hat{f}_m(j)
	&= \mathbb{E}_{z\sim U(\{ 0,1\}^n)}\left[\frac{1}{2^{n+m}} \left( 1 + \mu^{|z|}\right)^m (-1)^{j\cdot z} \right]\\
	&= \frac{1}{2^{n+m}} \sum\limits_{L=0}^m \binom{m}{L} \mathbb{E}_{z\sim U(\{ 0,1\}^n)}\left[\mu^{L|z|}(-1)^{j\cdot z} \right].
\end{align*}
We now rewrite the expectations on the right-hand side
\begin{align*}
	&\mathbb{E}_{z \sim U(\{ 0,1\}^n)}\left[\mu^{L|z|}(-1)^{j\cdot z} \right]\\
	&= \frac{1}{2^n}\sum\limits_{\ell=0}^n\sum\limits_{k=\max\{0,\ell-(n-|j|)\}}^{\min\{\ell,|j|\} }\binom{|j|}{k}\binom{n-|j|}{\ell - k}(-1)^k \mu^{L\cdot\ell}\\
	&= \frac{1}{2^n}\sum\limits_{k=0}^{|j|}\binom{|j|}{k}(-1)^k \sum\limits_{\ell=k}^{k+n-|j|} \binom{n-|j|}{\ell - k}\mu^{L\cdot\ell}\\
	&=\frac{1}{2^n}\sum\limits_{k=0}^{|j|}\binom{|j|}{k}(-1)^k \mu^{L\cdot k} \underbrace{\sum\limits_{\ell=0}^{n-|j|} \binom{n-|j|}{\ell}\mu^{L\cdot\ell}}_{=\left( 1+\mu^L\right)^{n-|j|}}\\
	&= \frac{\left( 1+\mu^L\right)^{n-|j|}}{2^n} \underbrace{\sum\limits_{k=0}^{|j|}\binom{|j|}{k}(-1)^k \mu^{L\cdot k}}_{=\left( 1-\mu^L\right)^{|j|}}\\
	&= \frac{\left( 1+\mu^L\right)^{n-|j|} \left( 1-\mu^L\right)^{|j|}}{2^n}.
\end{align*}
This allows us to upper bound the Fourier coefficients of $f$ as follows:
\begin{align*}
	\hat{f}_m(j)
	&= \frac{1}{2^{n+m}} \sum\limits_{L=0}^m \binom{m}{L} \left(\frac{1+\mu^L}{2} \right)^{n-|j|}\left(\frac{1-\mu^L}{2} \right)^{|j|}\\
	&\leq \frac{1}{2^{n+m}} \sum\limits_{L=0}^m \binom{m}{L} \left(\frac{1+\mu}{2} \right)^{n-|j|}\left(\frac{1-\mu^m}{2} \right)^{|j|}\\
	&= \frac{1}{2^{n}}\left(\frac{1+\mu}{2} \right)^{n-|j|}\left(\frac{1-\mu^m}{2} \right)^{|j|}.
\end{align*}
According to Lemma \ref{LmmDiagGramMatrix} this now gives us the following upper bound on the diagonal entries of the root of the Gram matrix
\begin{align*}
	\sqrt{G_m}(a,a)
	&\leq \frac{1}{2^n} \sum\limits_{j\in\{ 0,1\}^n}\sqrt{\left(\frac{1+\mu}{2} \right)^{n-|j|}\left(\frac{1-\mu^m}{2} \right)^{|j|}}\\
	&= \frac{1}{2^n} \sum\limits_{k=0}^n \binom{n}{k} \sqrt{\left(\frac{1+\mu}{2} \right)^{n-k}\left(\frac{1-\mu^m}{2} \right)^{k}}\\
	&= \frac{1}{2^n}\left( \sqrt{\frac{1+\mu}{2}} + \sqrt{\frac{1-\mu^m}{2}}\right)^n,
\end{align*}
and this in turn allows us to bound the PGM success probability as
\begin{align*}
	P^{PGM}(\mathcal{E})
	&= \sum\limits_{a\in\{ 0,1\}^n} \sqrt{G_m}(a,a)^2\\
	&\leq \frac{1}{2^n} \left( \sqrt{\frac{1+\mu}{2}} + \sqrt{\frac{1-\mu^m}{2}}\right)^{2n}\\
	&= \left(\frac{1}{2}\left(\sqrt{1+\mu} + \sqrt{1-\mu^m} \right) \right)^{2n}.
\end{align*}
We combine this with our learning requirement and Theorem \ref{ThmPGMSuccessProb} to obtain
\begin{align*}
	1-\delta \leq P^{opt}(\mathcal{E}) \leq \sqrt{P^{PGM}(\mathcal{E})} \leq \left(\frac{1}{2}\left(\sqrt{1+\mu} + \sqrt{1-\mu^m} \right) \right)^{n}.
\end{align*}
This can be rearranged (using $\delta <\frac{1}{3}$) to
\begin{align*}
	m
	&= \frac{-\log\left(1-\left(2\cdot\sqrt[n]{1-\delta}-\sqrt{1+\mu} \right)^2 \right)}{\log\frac{1}{\mu}}.
\end{align*}
With $\log(1+x)\leq x$ we obtain $\frac{1}{\log\frac{1}{\mu}}\geq \frac{1}{\frac{1}{\mu}-1}=\frac{\mu }{1-\mu}$ and $$-\log\left(1-\left(2\cdot\sqrt[n]{1-\delta}-\sqrt{1+\mu} \right)^2 \right)\geq \left(2\cdot\sqrt[n]{1-\delta}-\sqrt{1+\mu} \right)^2.$$
For $\mu\geq 1-\frac{1}{\ln(n)}$ we now obtain (for $n$ large enough)
\begin{align*}
	m
	&\geq (\ln (n)-1)\cdot \left(2\sqrt{\frac{2}{3}} - \sqrt{2} \right) = \Omega\left(\ln(n) \right),
\end{align*}
and this finishes the proof.
\end{proof}

Note that this proof strategy also yields for a strictly increasing function $g:\IN\to\IR_{>0}$ with $\lim\limits_{n\to\infty} g(n)=\infty$ and for a distribution $D_\mu$ with $\mu_i\geq 1-\frac{1}{g(n)}$ for all $1\leq i\leq n$ the sample complexity lower bound $\Omega (g(n))$ (for $n$ large enough). This is consistent with the intuition that solving the learner problem becomes harder when the distribution is more strongly biased towards the uninformative instance with all entries equal to $1$.\\

We now compare this lower bound to our previously obtained upper bounds.  First, we consider the $n$-independent part of the bounds. When comparing Theorem \ref{ThmAmpBiasBV1} with Lemma \ref{LmmQuSampLowerBoundDelta}, we obtain
\begin{align*}
\Omega\left(\frac{1}{c}\ln\left(\frac{1}{\delta}\right)\right)
\leq m
\leq \mathcal{O}\left(\left(\ln\left(\frac{1}{1-c+\frac{c^2}{2}}\right)\right)^{-1} \ln\left(\frac{1}{\delta}\right)\right).
\end{align*}
We study this for $\delta\ll 1$ (high confidence) and $c\ll 1$ (high bias). Then Taylor expansion shows
\begin{align*}
\left(\ln\left(\frac{1}{1-c+\frac{c^2}{2}}\right)\right)^{-1}
= \frac{1}{c} + \frac{c}{6} + \mathcal{O}(c^2)\quad\textrm{for } c\ll 1.
\end{align*}

Hence, lower and upper bound coincide in the relevant region for $\delta$ and $c$, so the $n$-independent part of the sample complexity upper bound provided by Algorithm \ref{AlgAmpBiasBV1} is optimal.\\

However, in comparing Theorem \ref{ThmAmpBiasBV2} with Lemma \ref{LmmQuSampLowerBoundDelta} we see a discrepancy between lower and upper bound for the relevant region $\delta \ll 1$ and $c-(1-\frac{1}{\sqrt{2n}})\ll 1$. Therefore we conjecture that the $c$-dependence of the upper bound arising from Theorem \ref{ThmAmpBiasBV2} is not optimal.\\

Now we compare the bounds w.r.t.~the $n$-dependence, i.e.~we compare Theorem \ref{ThmAmpBiasBV1} with Theorem \ref{ThmQuSampCompLowerBoundV2}, and obtain
\begin{align*}
\Omega\left(\ln(n)\right)
\leq m
\leq \mathcal{O}\left(\frac{1}{c}\ln(n)\right).
\end{align*}
But in Theorem \ref{ThmQuSampCompLowerBoundV2} we assumed that $\mu_i\geq 1 - \frac{1}{\ln (n)}$ for all $1\leq i\leq n$. When considering values for $\mu$ lying on this threshold we can rephrase this as condition on the (then $n$-dependent) c-boundedness parameter, namely $c\leq \frac{1}{\ln (n)}$. So when honestly including the $n$-dependence of $c$, our comparison becomes
\begin{align*}
\Omega\left(\ln(n)\right)
\leq m
\leq \mathcal{O}\left(\ln^2(n)\right)
\end{align*}
and is thus not tight.\\

\begin{figure}
\centering
\includegraphics[scale=0.5]{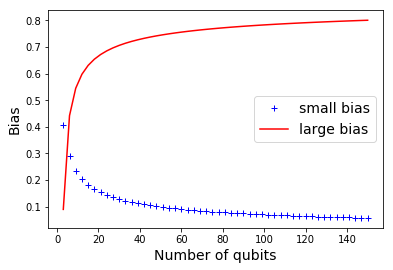}
\caption{A plot comparing the maximal bias allowed in Theorem \ref{ThmAmpBiasBV2} (depicted by the blue crosses) with the minimal bias required in Theorem \ref{ThmQuSampCompLowerBoundV2} (depicted by the red line).}
\label{Fig1}
\end{figure}

Finally, we want to point towards a second unsatisfactory aspect of our results. We provide an $n$-dependent quantum sample complexity lower bound for ``large'' noise and an $n$-independent quantum sample complexity upper bound for ``small'' noise. However, there is a large discrepancy between the obtained characterizations of ``small'' and ``large'' noise. That this already becomes relevant for moderate $n$ can be seen in figure \ref{Fig1}.\\

Hence, we did not succeed in identifying a bias threshold beyond which the sample complexity qualitatively differs from the unbiased case, but merely provided a region in which such a threshold would lie. To improve upon our results it would be necessary to modify either the proof of Theorem \ref{ThmAmpBiasBV2} to allow for stronger bias or the proof of Theorem \ref{ThmQuSampCompLowerBoundV2} to allow for weaker bias. In particular, it would be interesting to obtain a non-trivial quantum sample complexity lower bound for constant bias, i.e.~without introducing $n$-dependence into the $c$-boundedness parameter. However,  we currently do not see whether our proof strategies admit such an improvement.

\section{Conclusion and Outlook}
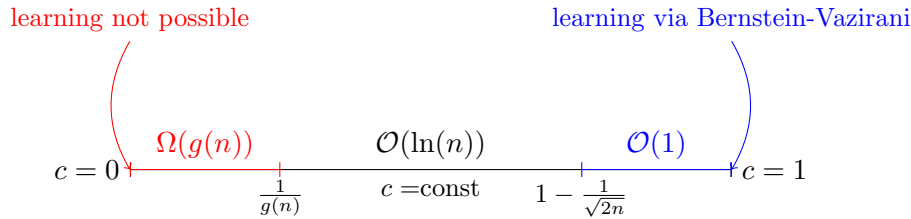
\begin{figure}
\begin{center}
\begin{tikzpicture}
\draw[|-|] (0,0) -- (8,0) node (1) at (0,0)[left]{$c=0$} node (2) at (8,0)[right]{$c=1$} node[midway,above] {$\mathcal{O}(\ln(n))$} node[midway,below] {\small $c=$const};
\node at (6,0)[below]{\small $1-\frac{1}{\sqrt{2n}}$};
\node at (2,0)[below]{\small $\frac{1}{g(n)}$};
\draw[|-|, red] (0,0) -- (2,0) node[midway,above] {$\Omega(g(n))$};
\draw[|-|, blue] (6,0) -- (8,0) node[midway,above] {$\mathcal{O}(1)$};
\node (T1) at (0,2){\textcolor{red}{\small learning not possible}};
\draw[->, red] (T1.south) to [out=240, in=120] (1.east);
\node (T2) at (8,2){\textcolor{blue}{\small learning via Bernstein-Vazirani}};
\draw[->, blue] (T2.south) to [out=300, in=60] (2.west);
\end{tikzpicture}
\caption{Overview of the quantum sample complexity upper and lower bounds from Theorems \ref{ThmAmpBiasBV1}, \ref{ThmAmpBiasBV2} and \ref{ThmQuSampCompLowerBoundV2} depending on the $c$-boundedness parameter (without noise in the training data). Here, $g:\IN\to\IR_{>0}$ is a strictly increasing function with $\lim\limits_{n\to\infty} g(n)=\infty$.}
\label{Fig2}
\end{center}
\end{figure}

In this paper we extended a well-known quantum learning strategy for linear functions from the uniform distribution to biased product distributions. This approach naturally led to a distinction between a procedure for arbitrary (not full) bias and a procedure for small bias, the latter with a significantly better performance. Moreover, we showed that the second procedure is (to a certain degree) stable w.r.t.~noise in the training data and in the performed quantum gates. Finally, we also provided lower bounds on the size of the training data required for the learning problem, both in the classical and in the quantum setting. The sample complexity upper and lower bounds in the case of no noise are summarized in figure \ref{Fig2}.\\

We want to conclude by outlining some open questions for future work:
\begin{itemize}
\item Can we identify a bias threshold s.t.~the optimal sample complexity below the threshold differs qualitatively from the one above it?
\item Is our learning procedure for small bias also stable w.r.t.~different types of noise in the training data, e.g.~malicious noise?
\item Our explicit learning algorithms also give upper bounds on the computational complexity of our learning problem. Can we find corresponding lower bounds to facilitate a discussion of optimality w.r.t.~runtime?
\item Can we find more examples of learning tasks (i.e.~function classes) where quantum training data yields an advantage w.r.t.~sample and/or time complexity?
\end{itemize}

\newpage
\section*{acknowledgements}
First, I want to thank my supervisor Michael Wolf for several stimulating discussions concerning questions of quantum learning. Also, I want to thank Benedikt Graswald for proofreading a first draft of this paper and for his constructive comments. Finally, I am grateful to Andrea Rocchetto for useful comments to improve the result of Appendix \ref{SbSctUnknownDistr} and for suggesting further references. Also, I thank the anonymous reviewers for their constructive criticism.\\
Support from the TopMath Graduate Center of the TUM Graduate School at the Technische Universität München, Germany, and from the TopMath Program at the Elite Network of Bavaria is gratefully acknowledged.

\bibliographystyle{plainnat}
\bibliography{Literature.bib}

\begin{thebibliography}{27}
\providecommand{\natexlab}[1]{#1}
\providecommand{\url}[1]{\texttt{#1}}
\expandafter\ifx\csname urlstyle\endcsname\relax
  \providecommand{\doi}[1]{doi: #1}\else
  \providecommand{\doi}{doi: \begingroup \urlstyle{rm}\Url}\fi

\bibitem[Arunachalam and de~Wolf(2017)]{Arunachalam.2017}
S.~Arunachalam and R.~de~Wolf.
\newblock Guest column: A survey of quantum learning theory.
\newblock \emph{SIGACT News}, 48, 2017.
\newblock \doi{10.1145/3106700.3106710}.
\newblock URL \url{https://pure.uva.nl/ws/files/25255496/p41_arunachalam.pdf}.

\bibitem[Arunachalam and de~Wolf(2018)]{Arunachalam.2018}
S.~Arunachalam and R.~de~Wolf.
\newblock Optimal quantum sample complexity of learning algorithms.
\newblock \emph{Journal of Machine Learning Research}, 19\penalty0
  (71):\penalty0 1--36, 2018.
\newblock ISSN 1533-7928.
\newblock URL \url{http://jmlr.org/papers/v19/18-195.html}.

\bibitem[Arunachalam et~al.(2018)Arunachalam, Chakraborty, Lee, Paraashar, and
  de~Wolf]{Arunachalam.20190307}
S.~Arunachalam, S.~Chakraborty, T.~Lee, M.~Paraashar, and R.~de~Wolf.
\newblock Two new results about quantum exact learning, 2018.
\newblock URL \url{https://arxiv.org/pdf/1810.00481}.

\bibitem[Arunachalam et~al.(2019)Arunachalam, Grilo, and
  Sundaram]{Arunachalam.20190307b}
S.~Arunachalam, A.~B. Grilo, and A.~Sundaram.
\newblock Quantum hardness of learning shallow classical circuits, 2019.
\newblock URL \url{https://arxiv.org/pdf/1903.02840}.

\bibitem[At{\i}c{\i} and Servedio(2007)]{Atc.2007}
A.~At{\i}c{\i} and R.~A. Servedio.
\newblock Quantum algorithms for learning and testing juntas.
\newblock \emph{Quantum Information Processing}, 6\penalty0 (5):\penalty0
  323--348, 2007.
\newblock ISSN 1570-0755.
\newblock \doi{10.1007/s11128-007-0061-6}.

\bibitem[Barnum and Knill(2000)]{Barnum.20000423}
H.~Barnum and E.~Knill.
\newblock Reversing quantum dynamics with near-optimal quantum and classical
  fidelity, 2000.
\newblock URL \url{http://arxiv.org/pdf/quant-ph/0004088}.

\bibitem[Benioff(1980)]{Benioff.1980}
P.~Benioff.
\newblock The computer as a physical system: A microscopic quantum mechanical
  hamiltonian model of computers as represented by turing machines.
\newblock \emph{Journal of Statistical Physics}, 22\penalty0 (5):\penalty0
  563--591, 1980.
\newblock ISSN 1572-9613.
\newblock \doi{10.1007/BF01011339}.

\bibitem[Bernstein and Vazirani(1993)]{Bernstein.1993}
E.~Bernstein and U.~Vazirani.
\newblock Quantum complexity theory.
\newblock In R.~Kosaraju, editor, \emph{Proceedings of the twenty-fifth annual
  ACM symposium on Theory of computing}, pages 11--20, New York, NY, 1993. ACM.
\newblock ISBN 0897915917.
\newblock \doi{10.1145/167088.167097}.

\bibitem[Blum et~al.(2003)Blum, Kalai, and Wasserman]{Blum.2003}
A.~Blum, A.~Kalai, and H.~Wasserman.
\newblock Noise-tolerant learning, the parity problem, and the statistical
  query model.
\newblock \emph{Journal of the ACM}, 50\penalty0 (4):\penalty0 506--519, 2003.
\newblock ISSN 00045411.
\newblock \doi{10.1145/792538.792543}.

\bibitem[Bshouty and Jackson(1998)]{Bshouty.1998}
N.~H. Bshouty and J.~C. Jackson.
\newblock Learning dnf over the uniform distribution using a quantum example
  oracle.
\newblock \emph{SIAM Journal on Computing}, 28\penalty0 (3):\penalty0
  1136--1153, 1998.
\newblock ISSN 0097-5397.
\newblock \doi{10.1137/S0097539795293123}.

\bibitem[Cross et~al.(2015)Cross, Smith, and Smolin]{Cross.2015}
A.~W. Cross, G.~Smith, and J.~A. Smolin.
\newblock Quantum learning robust against noise.
\newblock \emph{Physical Review A}, 92\penalty0 (1):\penalty0 97, 2015.
\newblock ISSN 1050-2947.
\newblock \doi{10.1103/PhysRevA.92.012327}.

\bibitem[Feynman(1985)]{Feynman.1985}
R.~P. Feynman.
\newblock Quantum mechanical computers.
\newblock \emph{Optics News}, 11\penalty0 (2):\penalty0 11, 1985.
\newblock ISSN 0098-907X.
\newblock \doi{10.1364/ON.11.2.000011}.

\bibitem[Grilo et~al.(2017)Grilo, Kerenidis, and Zijlstra]{Grilo.20180410}
Al.~B. Grilo, I.~Kerenidis, and T.~Zijlstra.
\newblock Learning with errors is easy with quantum samples, 2017.
\newblock URL \url{http://arxiv.org/pdf/1702.08255}.

\bibitem[Hausladen and Wootters(1994)]{Hausladen.1994}
P.~Hausladen and W.~K. Wootters.
\newblock A `pretty good' measurement for distinguishing quantum states.
\newblock \emph{Journal of Modern Optics}, 41\penalty0 (12):\penalty0
  2385--2390, 1994.
\newblock ISSN 0950-0340.
\newblock \doi{10.1080/09500349414552221}.

\bibitem[Hoeffding(1963)]{Hoeffding.1963}
W.~Hoeffding.
\newblock Probability inequalities for sums of bounded random variables.
\newblock \emph{Journal of the American Statistical Association}, 58\penalty0
  (301):\penalty0 13--30, 1963.
\newblock ISSN 0162-1459.
\newblock \doi{10.1080/01621459.1963.10500830}.

\bibitem[Ivanyos et~al.(2018)Ivanyos, Prakash, and Santha]{Ivanyos.2018}
G.~Ivanyos, A.~Prakash, and M.~Santha, editors.
\newblock \emph{On Learning Linear Functions from Subset and Its Applications
  in Quantum Computing: Schloss Dagstuhl - Leibniz-Zentrum fuer Informatik
  GmbH, Wadern/Saarbruecken, Germany}, 2018.
\newblock \doi{10.4230/LIPICS.ESA.2018.66}.

\bibitem[Kanade et~al.(2019)Kanade, Rocchetto, and Severini]{Kanade.2019}
V.~Kanade, A.~Rocchetto, and S.~Severini.
\newblock Learning dnfs under product distributions via {$\mu$}-biased quantum
  fourier sampling.
\newblock \emph{Quantum Information {\&} Computation}, 19\penalty0
  (15{\&}16):\penalty0 1261--1278, 2019.
\newblock URL
  \url{http://www.rintonpress.com/xxqic19/qic-19-1516/1261-1278.pdf}.

\bibitem[Lyubashevsky(2005)]{Lyubashevsky.2005}
V.~Lyubashevsky.
\newblock The parity problem in the presence of noise, decoding random linear
  codes, and the subset sum problem.
\newblock In C.~Chekuri, editor, \emph{Approximation, randomization and
  combinatorial optimization}, volume 3624 of \emph{Lecture Notes in Computer
  Science}, pages 378--389. Springer, Berlin, 2005.
\newblock ISBN 978-3-540-28239-6.
\newblock \doi{10.1007/11538462_32}.

\bibitem[Montanaro(2012)]{Montanaro.2012}
A.~Montanaro.
\newblock The quantum query complexity of learning multilinear polynomials.
\newblock \emph{Information Processing Letters}, 112\penalty0 (11):\penalty0
  438--442, 2012.
\newblock ISSN 00200190.
\newblock \doi{10.1016/j.ipl.2012.03.002}.

\bibitem[Nielsen and Chuang(2010)]{Nielsen.2010}
M.~A. Nielsen and I.~L. Chuang.
\newblock \emph{Quantum computation and quantum information}.
\newblock {Cambridge Univ. Press}, Cambridge, 10th anniversary ed. edition,
  2010.
\newblock ISBN 9781282967298.
\newblock URL
  \url{https://ebookcentral.proquest.com/lib/subhh/detail.action?docID=647366}.

\bibitem[O'Donnell(2014)]{ODonnell.2014}
R.~O'Donnell.
\newblock \emph{Analysis of boolean functions}.
\newblock {Cambridge University Press}, Cambridge, 2014.
\newblock ISBN 9781139814782.
\newblock URL \url{http://dx.doi.org/10.1017/CBO9781139814782}.

\bibitem[Regev(2009)]{Regev.2009}
O.~Regev.
\newblock On lattices, learning with errors, random linear codes, and
  cryptography.
\newblock \emph{Journal of the ACM}, 56\penalty0 (6):\penalty0 1--40, 2009.
\newblock ISSN 00045411.
\newblock \doi{10.1145/1568318.1568324}.

\bibitem[Rist{\`e} et~al.(2017)Rist{\`e}, {da Silva}, Ryan, Cross,
  C{\'o}rcoles, Smolin, Gambetta, Chow, and Johnson]{Riste.2017}
D.~Rist{\`e}, M.~P. {da Silva}, C.~A. Ryan, A.~W. Cross, A.~D. C{\'o}rcoles,
  J.~A. Smolin, J.~M. Gambetta, J.~M. Chow, and B.~R. Johnson.
\newblock Demonstration of quantum advantage in machine learning.
\newblock \emph{npj Quantum Information}, 3\penalty0 (1):\penalty0 16, 2017.
\newblock ISSN 2056-6387.
\newblock \doi{10.1038/s41534-017-0017-3}.

\bibitem[Servedio and Gortler(2004)]{Servedio.2004}
R.~A. Servedio and S.~J. Gortler.
\newblock Equivalences and separations between quantum and classical
  learnability.
\newblock \emph{SIAM Journal on Computing}, 33\penalty0 (5):\penalty0
  1067--1092, 2004.
\newblock ISSN 0097-5397.
\newblock \doi{10.1137/S0097539704412910}.

\bibitem[Shalev-Shwartz and Ben-David(2014)]{ShalevShwartz.2014}
S.~Shalev-Shwartz and S.~Ben-David.
\newblock \emph{Understanding machine learning: From theory to algorithms}.
\newblock {Cambridge University Press}, Cambridge, 2014.
\newblock ISBN 9781107057135.
\newblock \doi{10.1017/CBO9781107298019}.
\newblock URL \url{https://doi.org/10.1017/CBO9781107298019}.

\bibitem[Valiant(1984)]{Valiant.1984}
L.~G. Valiant.
\newblock A theory of the learnable.
\newblock \emph{Communications of the ACM}, 27\penalty0 (11):\penalty0
  1134--1142, 1984.
\newblock ISSN 00010782.
\newblock \doi{10.1145/1968.1972}.

\bibitem[Vershynin(2018)]{Vershynin.2018}
R.~Vershynin.
\newblock \emph{High-dimensional probability: An introduction with applications
  in data science}, volume~47 of \emph{Cambridge series in statistical and
  probabilistic mathematics}.
\newblock {Cambridge University Press}, Cambridge, 2018.
\newblock ISBN 9781108231596.
\newblock URL \url{https://doi.org/10.1017/9781108231596}.

\end{thebibliography}

\newpage
\appendix
\section*{Appendix}
\setcounter{secnumdepth}{2}

\section{Stability w.r.t.~Noise}\label{SctNoiseStability}
Both algorithms presented in section \ref{SctQuantSampCompUpperBounds} implicitly assume that the quantum example state perfectly represents the underlying function and that all quantum gates performed during the computation are perfectly accurate. In this section we relax these assumptions. We will do so separately, but our analysis shows that moderate noise in the training data and moderately faulty quantum gates can be tolerated at the same time.

\subsection{Noisy Training Data}
One of the most well-studied noise models in classical learning theory is that of random classification noise. Here, the training data is assumed to be s.t.~with probability $1-\eta$, the learning algorithm obtains a correct example, and with probability $\eta$, the examples label is flipped. In \cite{Arunachalam.2017}, this is translated to a quantum example state which in our notation has the form
\begin{align*}
|\varphi_a^{\textrm{noisy}}\rangle = 
~ \sqrt{1-\eta}&\left(\sum\limits_{x\in\{ -1,1\} } \sqrt{D_\mu (x)}|x,f^{(a)}(x)\rangle\right)\\
+ \sqrt{\eta}&\left(\sum\limits_{x\in\{ -1,1\} } \sqrt{D_\mu (x)}|x,f^{(a)}(x)\oplus 1\rangle\right).
\end{align*}
We will only shortly comment on how to battle this type of noise with our learning strategy at the end of this subsection. Instead, our focus will be on a performance analysis of our algorithm in the case of noisy training data similar to \cite{Grilo.20180410}. This means that we now assume our quantum example state to be of the form
\begin{align*}
|\psi_{a}^{\textrm{noisy}}\rangle =\sum_{x\in\lbrace-1,1\rbrace^{n}}\sqrt{D_{\mu}(x)}|x,\sum\limits_{i=1}^n a_i\frac{1-x_i}{2} + \xi^i_{x_i}\rangle,
\end{align*}
where the $\xi^i_{x_i}$, for $1\leq i\leq n$ and $x_i\in\lbrace -1,1\rbrace$, are independent random variables distributed according to Bernoulli distributions with parameters $\eta^i$ (i.e.~$\mathbb{P}[\xi^i_{x_i} = 1] = \eta^i = 1-\mathbb{P}[\xi^i_{x_i} = 0]$ for all $1\leq i\leq n$) and addition is understood modulo $2$.\\

Here, we choose a noise model that is rather general but we make an important restriction. Namely, we do not allow a noise $\xi_x$ that depends in an arbitrary way on $x$ but rather we require the noise to have a specific sum structure $\xi_x = \sum\limits_{i=1}^n \xi^i_{x_i}$. This requirement will later imply that also the noisy Fourier coefficients factorise. As this factorization is crucial for our analysis, with our strategy we cannot generalize the results of \cite{Grilo.20180410} on that more general noise model.\\

We first examine the result of applying the same procedure as in Algorithm \ref{AlgBiasBV} to a copy of a noisy quantum example state $|\psi_{a}^{\textrm{noisy}}\rangle$. To simplify referencing we write this down one more time as Algorithm \ref{AlgNoisyBiasBV} even though the procedure is exactly the same, only the form of the input changes.

\begin{algorithm}
\caption{Generalised Bernstein-Vazirani algorithm with noisy training data}\label{AlgNoisyBiasBV}
\begin{flushleft}
\hspace*{\algorithmicindent} \textbf{Input}: $|\psi_{a}^{\textrm{noisy}}\rangle =\sum\limits_{x\in\lbrace-1,1\rbrace^{n}}\sqrt{D_{\mu}(x)}|x,\sum\limits_{i=1}^n a_i\frac{1-x_i}{2} + \xi^i_{x_i}\rangle$, as well as $\mu\in [-1,1]$\\
\hspace*{\algorithmicindent} \textbf{Output}: See Theorem \ref{ThmBiasBVNoisyProb}\\
\hspace*{\algorithmicindent} \textbf{Success Probability}: $\frac{1}{2}$.
\end{flushleft}
\begin{algorithmic}[1]
\State Perform the $\mu$-biased QFT $H_\mu$ on the first $n$ qubits, obtain the state $(H_\mu \otimes \mathds{1})|\psi_{a}^{\textrm{noisy}}\rangle$.
\State Perform a Hadamard gate on the last qubit, obtain the state $(H_\mu \otimes H)|\psi_{a}^{\textrm{noisy}}\rangle$.
\State Measure each qubit in the computational basis and observe outcome $j = j_1\ldots j_{n+1}$.
\If{$j_{n+1} = 0$}\Comment{This corresponds to a failure of the algorithm.}
	\State Output $o = \perp$. 
\ElsIf{$j_{n+1} = 1$}\Comment{This corresponds to a success of the algorithm.}
	\State Output $o = j_1\ldots j_n$.
\EndIf
\end{algorithmic}
\end{algorithm}

Similarly to our previous analysis, we will first study the Fourier coefficients that are relevant for the sampling process in Algorithm \ref{AlgNoisyBiasBV}.

\begin{lemma}\label{LmmBiasedFourierCoeffNoisy}
Let $a\in\lbrace 0,1\rbrace^n$. Let $\xi^i_{x_i}$, for $1\leq i\leq n$ and $x_i\in\lbrace -1,1\rbrace$, be independent Bernoulli distributions, let  $g^{(a)}(x):=(-1)^{\sum\limits_{i=1}^n a_i\frac{1-x_i}{2} + \xi^i_{x_i}}$ and let $\mu\in (-1,1)$. Then the $\mu$-biased Fourier coefficients of $g^{(a)}$ satisfy: For $y\in\lbrace 0,1\rbrace^n$, with probability \[\prod\limits_{l=1}^n \left(y_l \cdot 2\eta^l (1-\eta^l) + (1-y_l)\cdot (1-2\eta^l (1-\eta^l))\right),\] it holds that 
\begin{align*}
\hat{g}_\mu^{(a)}(j) = &\prod\limits_{l:a_l=0}\left(y_l\cdot (-1)^{b_l}\left((1-j_l)\mu_l + j_l\sqrt{1-\mu_l^2}\right) + (1-y_l)\cdot (-1)^{b_l}(1-j_l)\right)\\
\cdot &\prod\limits_{l:a_l=1}\left(y_l\cdot (-1)^{b_l}(1-j_l) + (1-y_l)\cdot (-1)^{b_l}\left((1-j_l)\mu_l + j_l\sqrt{1-\mu_l^2}\right)\right).
\end{align*}
\end{lemma}

\begin{proof}
The proof is analogous to the one of Lemma \ref{LmmBiasedFourierCoeff}, see Appendix \ref{SctProofs}.
\end{proof}

We now make a step analogous to the one from Lemma \ref{LmmBiasedFourierCoeff} to Theorem \ref{ThmBiasBVProb} in order to understand the output of Algorithm \ref{AlgNoisyBiasBV}.

\begin{theorem}\label{ThmBiasBVNoisyProb}
Let $|\psi_{a}^{\textrm{noisy}}\rangle =\sum\limits_{x\in\lbrace-1,1\rbrace^{n}}\sqrt{D_{\mu}(x)}|x,\sum\limits_{i=1}^n a_i\frac{1-x_i}{2} + \xi^i_{x_i}\rangle$ be a noisy quantum example state, $a\in\lbrace 0,1\rbrace^n$, $\mu\in (-1,1)^n$. Then Algorithm \ref{AlgNoisyBiasBV} provides an outcome $|j_1\ldots j_{n+1}\rangle$ with the following properties:
\begin{enumerate}[(i)]
\item $\mathbb{P}[j_{n+1}=0]=\frac{1}{2}=\mathbb{P}[j_{n+1}=1].$
\item For any $1\leq i\leq n$, with probability $1-2\eta^i(1-\eta^i)$ it holds that
\begin{align*}
\mathbb{P}[a_i &= 0\neq j_i|j_{n+1}=1] = 0,\quad
\mathbb{P}[a_i = 1\neq j_i|j_{n+1}=1] = \mu^2.
\end{align*}
\item For any $1\leq i\leq n$, with probability $2\eta^i(1-\eta^i)$ it holds that
\begin{align*}
\mathbb{P}[a_i &= 0\neq j_i|j_{n+1}=1] = 1-\mu^2,\quad
\mathbb{P}[a_i = 1\neq j_i|j_{n+1}=1] = 1.
\end{align*}
\end{enumerate}
\end{theorem}

Note that in the scenario of Theorem \ref{ThmBiasBVNoisyProb} the underlying distribution $D_\mu$ is known to the algorithm as $\mu$ is provided as part of the input (see Algorithm \ref{AlgNoisyBiasBV}). Building on this subroutine we will now describe an amplified procedure for moderate noise (which is made precise in Theorem \ref{ThmAmpBiasBVNoisy}) in Algorithm \ref{AlgNoisyAmpBV} analogous to the one described in subsection $5.2$. Again, only the input changes, but we write the procedure down explicitly to simplify referencing.

\begin{algorithm}
\caption{Amplified Generalised Bernstein-Vazirani algorithm with noisy training data}\label{AlgNoisyAmpBV}
\begin{flushleft}
\hspace*{\algorithmicindent} \textbf{Input}: $m$ copies of $|\psi_{a}^{\textrm{noisy}}\rangle =\sum\limits_{x\in\lbrace-1,1\rbrace^{n}}\sqrt{D_{\mu}(x)}|x,\sum\limits_{i=1}^n a_i\frac{1-x_i}{2} + \xi^i_{x_i}\rangle$ for $a\in\lbrace0,1\rbrace^{n}$, where the number of copies is  $m\geq C\left(\max\left\lbrace \frac{1}{(1-5n\rho)^2},\frac{1}{(1-4n(1-c)^2)^2}\right\rbrace\ln\left(\frac{1}{\delta}\right)\right)$, as well as $\mu\in [-1,1]^n$ and $c\in (0,1]$ s.t.~$D_\mu$ is $c$-bounded.\\
\hspace*{\algorithmicindent} \textbf{Output}: $a\in\lbrace0,1\rbrace^{n}$\\
\hspace*{\algorithmicindent} \textbf{Success Probability}: $\geq 1-\delta$
\end{flushleft}
\begin{algorithmic}[1]
\For{$1\leq l \leq m$}
	\State Run Algorithm \ref{AlgNoisyBiasBV} on the $l^{th}$ copy of $|\psi_{a}^{\textrm{noisy}}\rangle$, store the output as $o^{(l)}$.
\EndFor
\If{$\exists 1\leq l\leq m: o^{(l)}\neq\perp$}
	\For{$1\leq i\leq n$}
		\State Let $o_i = \argmax\limits_{r\in\lbrace 0,1\rbrace} \lvert\lbrace 1\leq l\leq m | o^{(l)}_i = r\rbrace\rvert$.
	\EndFor
	\State Output $o=o_1 \ldots o_n$.
\ElsIf{$\forall 1\leq l\leq m: o^{(l)}=\perp$}
	\State Output $o=\perp$.
\EndIf
\end{algorithmic}
\end{algorithm}

\begin{theorem}\label{ThmAmpBiasBVNoisy}
Let $|\psi_{a}^{\textrm{noisy}}\rangle =\sum\limits_{x\in\lbrace-1,1\rbrace^{n}}\sqrt{D_{\mu}(x)}|x,\sum\limits_{i=1}^n a_i\frac{1-x_i}{2} + \xi^i_{x_i}\rangle$, with $a\in\lbrace 0,1\rbrace^n$, $\mu\in (-1,1)^n$ s.t.~$D_\mu$ is $c$-bounded for some $c\in (0,1]$ satisfying $c>1-\frac{1}{2\sqrt{n}}$. Further assume that $2\eta^i(1-\eta^i)<\frac{1}{5n}$ for all $1\leq i\leq n$, write $\rho:=\max\limits_{1\leq i\leq n}2\eta^i(1-\eta^i)$ . Then $\mathcal{O}\left(\max\left\lbrace \frac{1}{(1-5n\rho)^2},\frac{1}{(1-4n(1-c)^2)^2}\right\rbrace\ln\left(\frac{1}{\delta}\right)\right)$ copies of the quantum example state $|\psi_{a}\rangle$ suffice to guarantee that with probability $\geq 1-\delta$ Algorithm \ref{AlgNoisyAmpBV} outputs the string $a$.
\end{theorem}

As in Theorem \ref{ThmAmpBiasBV2}, our restrictions on both the $c$-boundedness parameter and the noise strength lead to a basically $n$-independent sample complexity upper bound.\\

\begin{proof}
The proof is analogous to the one of Theorem \ref{ThmAmpBiasBV2}, see Appendix \ref{SctProofs}.
\end{proof}

The previous Theorem shows that if the bias is not too strong and if the noise is not too random (i.e.~the probability of adding a random $1$ is either very low or very high), then learning is possible with essentially the same sample complexity as in the case without noise (compare Theorem \ref{ThmAmpBiasBV2}).\\
Note that the proof of Theorem \ref{ThmAmpBiasBVNoisy} shows that the exact choices of the bounds (in our formulation $c>1-\frac{1}{2\sqrt{n}}$ and $2\eta^i(1-\eta^i)<\frac{1}{5n}$) are flexible to some degree with a trade-off. If we have a better bound on $c$, we can loosen our requirement on the $\eta^i$ and vice versa.\\

Also observe that the requirement of ``not too random noise'' is natural. If $2\eta^i(1-\eta^i)\to \frac{1}{2}$ or, equivalently, $\eta^i\to\frac{1}{2}$, then the label in the noisy quantum example state becomes completely random and thus no information on the string $a$ can be extracted from it. Our bound gives a quantitative version of this intuition.\\
Nevertheless, the restriction which we put on the noise can be considered quite strong because of its $n$-dependence. This can, however, be relaxed at the cost of a looser sample complexity upper bound. Namely, similarly to the difference between the proofs of Theorem \ref{ThmAmpBiasBV1} and \ref{ThmAmpBiasBV2}, if we e.g.~only assume $2\eta^i(1-\eta^i)<\frac{1}{5}$ for all $1\leq i\leq n$, we can first for each coordinate separately bound the probability of the noise variables becoming relevant in at least $\frac{k}{5}$ runs using Hoeffding's inequality and then use the union bound. This will yield a quantum sample complexity upper bound with an $n$-dependent term of the form $\ln (n)$. Hence, if we assume a $c$-boundedness parameter strongly restricted as in Theorems \ref{ThmAmpBiasBV2} or \ref{ThmAmpBiasBVNoisy}, but obtain faulty training data states without an $n$-dependent noise bound as in Theorem \ref{ThmAmpBiasBVNoisy}, then we can still obtain a sample complexity upper bound with the same $n$-dependence as in Theorem \ref{ThmAmpBiasBV1}.\\

Finally, as promised at the beginning of this subsection, we shortly describe how to use the ideas presented in this subsection in the case of random classification noise as in \cite{Arunachalam.2017}. If the quantum learning algorithm has access to copies of a quantum example state
\begin{align*}
|\varphi_a^{\textrm{noisy}}\rangle
= \sqrt{1-\eta}&\left(\sum\limits_{x\in\{ -1,1\} } \sqrt{D_\mu (x)}|x,f^{(a)}(x)\rangle\right)\\ + \sqrt{\eta}&\left(\sum\limits_{x\in\{ -1,1\} } \sqrt{D_\mu (x)}|x,f^{(a)}(x)\oplus 1\rangle\right),
\end{align*}
then we observe that applying the $\mu$-biased Fourier transform to the first $n$ qubits and the standard Fourier transform to the last qubit gives
\begin{align*}
(H_\mu^{\otimes n}\otimes H)\left(|\varphi_a^{\textrm{noisy}}\rangle\right)
&= \frac{\sqrt{1-\eta}+\sqrt{\eta}}{\sqrt{2}}|0,\ldots,0\rangle \\
&\hphantom{=}+ \frac{\sqrt{1-\eta}-\sqrt{\eta}}{\sqrt{2}}\sum\limits_{j\in\{ 0,1\}} \hat{g}_\mu (j) |j,1\rangle.
\end{align*}
Hence, compared to the scenario studied in section \ref{SctQuantSampCompUpperBounds} the probabilities of observing a certain string as measurement outcome are simply scaled by a factor of $(\sqrt{1-\eta}\pm\sqrt{\eta})^2=1\pm 2\sqrt{\eta (1-\eta)}$. So our analysis carries over almost directly. We do not give the detailed reasoning here but only mention that incorporating the now rescaled probabilities basically changes the sample complexity upper bounds from the non-noisy case by a factor of $\frac{1}{(\eta - \frac{1}{2})^2}$, which is again in accordance with the intuition that the learning task becomes hard - and eventually impossible - for $\eta\to\frac{1}{2}$.

\subsection{Faulty Quantum Gates}
We now turn to the (more realistic) setting where the quantum gates in our computation (i.e.~the $\mu$-biased quantum Fourier transforms) are not implemented exactly but only approximately. In thi scenario, we obtain

\begin{lemma}\label{LmmAmpBiasBVNoisyGatesProb}
Let $|\psi_{a}\rangle =\sum\limits_{x\in\lbrace-1,1\rbrace^{n}}\sqrt{D_{\mu}(x)}|x,f^{(a)}(x)\rangle$ be a quantum example state, with $a\in\lbrace 0,1\rbrace^n$, $\mu\in (-1,1)^n$. Then a version of Algorithm \ref{AlgBiasBV} with $H_\mu$ replaced by $H_{\tilde{\mu}}$ for $\norm{H_{\mu} - H_{\tilde{\mu}}}_2 \leq\varepsilon$ provides an outcome $|j_1\ldots j_{n+1}\rangle$ with the following properties:
\begin{enumerate}[(i)]
\item $|\mathbb{P}[j_{n+1}=0] - \frac{1}{2}|\leq\varepsilon$ and $|\mathbb{P}[j_{n+1}=1]- \frac{1}{2}|\leq\varepsilon$,
\item $|\mathbb{P}[j_1\ldots j_n = a|j_{n+1}=1] - \prod\limits_{l:a_l=1} (1-\mu_l^2)|\leq\varepsilon$,
\item for $c\neq a$: $$\left\lvert\mathbb{P}[j_1\ldots j_n = c|j_{n+1}=1] - \prod\limits_{l:a_l=0} (1-c_l)\cdot \prod\limits_{l:a_l=1} \left( (1-c_l)\mu_l^2 + c_l (1-\mu_l^2) \right)\right\rvert\leq\varepsilon,$$
\item $\mathbb{P}[\exists 1\leq i\leq n: a_i = 0 \neq j_i | j_{n+1} = 1] \leq\varepsilon$, and
\item $\mathbb{P}[\exists 1\leq i\leq n: a_i = 1 \neq j_i | j_{n+1} = 1] \leq\sum\limits_{i=1}^n \mu_i^2 + \varepsilon$. In particular, if $D_\mu$ is $c$-bounded, then $\mathbb{P}[\exists 1\leq i\leq n: a_i = 1 \neq j_i | j_{n+1} = 1] \leq n(1-c)^2 + \varepsilon$.
\end{enumerate}
\end{lemma}
\begin{proof}
This follows from Theorem \ref{ThmBiasBVProb} because the outcome probabilities are the squares of the amplitudes and thus the difference in outcome probabilities can be bounded by the $2$-norm of the difference of the quantum states after applying the biased quantum Fourier transform and its approximate version.
\end{proof}

Now we can proceed analogously to the proof strategy employed in Theorem \ref{ThmAmpBiasBVNoisy} to derive

\begin{theorem}\label{ThmAmpBiasBVNoisyGates}
Let $|\psi_{a}\rangle =\sum\limits_{x\in\lbrace-1,1\rbrace^{n}}\sqrt{D_{\mu}(x)}|x,f^{(a)}(x)\rangle$, $a\in\lbrace 0,1\rbrace^n$, $\mu\in (-1,1)^n$ s.t.~$D_\mu$ is $c$-bounded for some $c\in (0,1]$ satisfying $c >1-\sqrt{\frac{1-2\varepsilon}{2n}}$. Then $$\mathcal{O}\left(\max\left\lbrace \frac{1}{(1-2\varepsilon)^2},\frac{1}{1-2(n(1-c)^2+\varepsilon)^2}\right\rbrace\ln\left(\frac{1}{\delta}\right) + \varepsilon\right)$$ copies of the quantum example state $|\psi_{a}\rangle$ suffice to guarantee that, with probability $\geq 1-\delta$, a version of Algorithm \ref{AlgAmpBiasBV2} with $H_\mu$ replaced by $H_{\tilde{\mu}}$ for $\norm{H_{\mu} - H_{\tilde{\mu}}}_2 \leq\varepsilon\in (0,\frac{1}{2})$ outputs the string $a$.
\end{theorem}

In particular, the sample complexity upper bound from Theorem \ref{ThmAmpBiasBV2} remains basically untouched if quantum gates with small error are used.

\subsection{The Case of Unknown Underlying Distributions}\label{SbSctUnknownDistr}
An interesting consequence of the result of the previous subsection is the possibility to drop the assumption of prior knowledge of the underlying product distribution, as was already observed in \cite{Kanade.2019} for a similar scenario. The important observations towards this end are given in this subsection.

\begin{lemma}\label{LmmApproxProdUnitaries}\emph{(Lemma $5$ in \cite{Kanade.2019})}\\
Let $A=A_n\cdots A_1$ be a product of unitary operators $A_j$. Assume that for every $A_j$ there exists an approximation $\tilde{A}_j$ s.t.~$\norm{A_j-\tilde{A}_j}\leq\varepsilon_j$. Then it holds that
\begin{align*}
\norm{A_n\cdots A_1 - \tilde{A}_n\cdots \tilde{A}_1}\leq\sum\limits_{j=1}^n\varepsilon_j,
\end{align*}
i.e.~the operator $\tilde{A}:=\tilde{A}_n\cdots \tilde{A}_1$ is an $\varepsilon$-approximation to $A$ w.r.t.~the operator norm.
\end{lemma}
\begin{proof}
This can be proven by induction using the triangle inequality and the fact that a unitary operator has operator norm equal to $1$. For details, the reader is referred to \cite{Kanade.2019}.
\end{proof}

This can be used to derive (compare again \cite{Kanade.2019})
\begin{corollary}\label{CrlApproxBiasQuFourierTrafo}
Let $\mu\in (-1,1)^n$ be s.t.~the distribution $D_\mu$ is $c$-bounded for $c\in (0,1]$. Let $\tilde{\mu}\in (-1,1)^n$ satisfy $\norm{\mu-\tilde{\mu}}_\infty\leq\varepsilon$. Then the corresponding biased quantum Fourier transforms satisfy
\begin{align*}
\norm{H_\mu - H_{\tilde{\mu}}}\leq 2\sqrt{2} n\gamma\varepsilon,
\end{align*}
where $\gamma = \frac{1}{c^2}\left( (2-c)\frac{3}{2\sqrt{2}c} + 1\right)$.
\end{corollary}
\begin{proof}
This proof is given in Appendix \ref{SctProofs}.
\end{proof}

The next Lemma is on approximating the bias parameter of an unknown product distribution from examples. (Compare the closing remark in Appendix A of \cite{Kanade.2019}.)

\begin{lemma}\label{LmmApproximatingBiasParam}
Using $m\leq\mathcal{O}(\frac{8\gamma^2 \cdot n^2}{\varepsilon^2}\ln(\frac{n}{\delta}))$ copies of the quantum example state $|\psi_{a}\rangle$ (or of $|\psi_{a}^{\textrm{noisy}}\rangle$) for a product distribution $D_\mu$ with bias vector $\mu\in (-1,1)^n$ s.t.~$D_\mu$ is $c$-bounded for $c\in (0,1]$ one can, with probability $\geq 1-\delta$, output $\tilde{\mu}\in (-1,1)^n$ s.t.~$\norm{H_\mu - H_{\tilde{\mu}}}\leq\varepsilon$.
\end{lemma}
\begin{proof}
Recall that $\mu_i = \mathbb{E}_{D_\mu}[x_i]$. Via a standard application of Hoeffding's inequality we conclude that $\mathcal{O}(\frac{8\gamma^2 \cdot n^2}{\varepsilon^2} \ln(\frac{1}{\delta}))$ examples drawn i.i.d.~from $D_\mu$ (which can be obtained from copies of the quantum example state by measuring the corresponding subsystem) are sufficient to guarantee that, with probability $\geq 1-\delta$, the empirical estimate $\hat{\mu}_i$ satisfies $|\mu_i - \hat{\mu}_i|\leq \frac{\varepsilon}{2\sqrt{2}\gamma \cdot n}$. 
As each component of a copy of the quantum example state can be measured separately, we see - using the union bound, that $\mathcal{O}(\frac{8\gamma^2 \cdot n^2}{\varepsilon^2}\ln(\frac{n}{\delta}))$ copies of the (possibly noisy) quantum example state suffice to guarantee that, with probability $\geq 1-\delta$, it holds that $\norm{\mu-\hat{\mu}}_\infty \leq \frac{\varepsilon}{2\sqrt{2}\gamma \cdot n}$. Now we can apply the previous Corollary to finish the proof.
\end{proof}

If we now combine this result with Theorem \ref{ThmAmpBiasBVNoisyGates}, we obtain a sample complexity upper bound for our learning problem without assuming the underlying distribution to be known in advance.

\begin{corollary}\label{CrlQuLearnLinUnknownDistr}
Let $|\psi_{a}\rangle =\sum\limits_{x\in\lbrace-1,1\rbrace^{n}}\sqrt{D_{\mu}(x)}|x,f^{(a)}(x)\rangle$, $a\in\lbrace 0,1\rbrace^n$, $\mu\in (-1,1)^n$ s.t.~$D_\mu$ is $c$-bounded for some $c\in (0,1]$ satisfying $c >1-\sqrt{\frac{1-2\varepsilon}{2n}}$. Then there exists a quantum algorithm which, given access to $$\mathcal{O}\left(\frac{8\gamma^2 \cdot n^2}{\varepsilon^2}\ln(\frac{n}{\delta}) + \max\left\lbrace \frac{1}{(1-2\varepsilon)^2},\frac{1}{1-2(n(1-c)^2+\varepsilon)^2}\right\rbrace\ln\left(\frac{1}{\delta}\right)\right)$$ copies of the quantum example state $|\psi_{a}\rangle$, with probability $\geq 1-\delta$, outputs the string $a$, without prior knowledge of the underlying distribution $D_\mu$.
\end{corollary}

Note, however, that the learning algorithm does need to obtain the $c$-boundedness parameter $c$ as input in advance, but this (in general) does not fix the underlying distribution.\\
Observe also that - since Lemma \ref{LmmApproximatingBiasParam} remains valid for noisy quantum examples -, even though we do not explicitly formulate the result of this subsection for noisy quantum training data, such a generalization is possible by combining the strategies presented in this and the previous subsections.

\section{Proofs}\label{SctProofs}
\noindent\textbf{Proof of Lemma \ref{LmmBiasedQFourierSampling}:}\\
We directly compute the state produced by the algorithm before the measurement is performed:
\begin{align*}
	(H_\mu \otimes H)|\psi_{f}\rangle
	&= \sum\limits_{x\in\{ -1,1\}^n}\sum\limits_{j\in \{ 0,1\}^n} \frac{1}{\sqrt{2}} D_\mu (x) \phi_{\mu,j}(x)\left(|j,0\rangle + (-1)^{f(x)}|j,1\rangle \right)\\
	&= \frac{1}{\sqrt{2}}\sum\limits_{j\in \{ 0,1\}^n} \underbrace{\mathbb{E}_{D_\mu}[\phi_{\mu,j}]}_{=\delta_{j,0\ldots 0}}|j,0\rangle + \underbrace{\mathbb{E}_{D_\mu}[g\phi_{\mu,j}]}_{=\hat{g}_\mu(j)}|j,1\rangle.
\end{align*}
Hence, the computational basis measurement from step $3$ of Algorithm \ref{AlgBiasQFS} on the last qubit returns $1$ with probability $\frac{1}{2}$ and if that is the case, the computational basis measurement on the first $n$ qubits will return $j$ with probability $\left(\hat{g}_{\mu}(j)\right)^{2}$, as claimed. \hfill $\square$\\

\noindent\textbf{Proof of Lemma \ref{LmmDiagGramMatrix}:}\\
The proof is by direct computation using the Fourier expansion:
\begin{align*}
	(H G H^{-1})(a,b) 
	&= \frac{1}{2^n}\sum\limits_{c,d\in\{ 0,1\}^n} (-1)^{c\cdot a + d\cdot b} g(c+d)\\
	&= \frac{1}{2^n}\sum\limits_{c,d,j\in\{ 0,1\}^n} (-1)^{c\cdot a + d\cdot b + j\cdot(c+d)}\hat{g}(j)\\
	&= \frac{1}{2^n}\sum\limits_{j\in\{ 0,1\}^n}\hat{g}(j)\underbrace{\sum\limits_{c\in\{ 0,1\}^n}(-1)^{c\cdot (a+j)}}_{=2^n\delta_{a,j}} \underbrace{\sum\limits_{d\in\{ 0,1\}^n}(-1)^{d\cdot (b+j)}}_{=2^n\delta_{b,j}}\\
	&= 2^n\hat{g}(a)\delta_{a,b}.
\end{align*}
Unitarity of $H$ can be checked easily by exploiting the same identity as in the second to last line of the previous computation. \hfill $\square$\\

\noindent\textbf{Proof of Corollary \ref{CrlDiagEntriesRootGram}:}\\
Using Lemma \ref{LmmDiagGramMatrix} we can directly compute the diagonal entries of the matrix root and obtain
\begin{align*}
	\sqrt{G}(a,a) 
	&= \left(H^{-1}\cdot \textrm{diag}\left(\left\lbrace \sqrt{2^n \hat{g}(j)}~|~ j\in\{ 0,1\}^n\right\rbrace\right)\cdot H\right)(a,a)\\
	&= \frac{1}{2^n}\sum\limits_{j,k\in\{ 0,1\}^n} (-1)^{c\cdot j + d\cdot k}\sqrt{2^n \hat{g}(j)}\delta_{j,k}\\
	&= \frac{1}{\sqrt{2^n}}\sum\limits_{j\in\{ 0,1\}^n} \sqrt{\hat{g}(j)}
\end{align*}
for every $a\in\{ 0,1\}^n$.\hfill $\square$\\

\noindent\textbf{Proof of Lemma \ref{LmmBiasedFourierCoeffNoisy}:}\\
As in the proof of Lemma \ref{LmmBiasedFourierCoeff}, due to the product structure of all the relevant objects (here our assumption on the form of the noise enters), it suffices to consider the case $n=1$ in detail. In this case we have $f^{(a)}(x)=a\tilde{x}$, $g^{(a)}(x)=(-1)^{a\tilde{x} + \xi_x}$ for $\tilde{x}=\frac{1-x}{2}$, $\phi_{\mu,0}(x)=1$, and $\phi_{\mu,1}(x)=\frac{x-\mu}{\sqrt{1-\mu^2}}$. (We leave out unnecessary indices to improve readability.) We compute
\begin{align*}
\hat{g}_\mu^{(a)}(j) 
&= \mathbb{E}_{D_\mu}[(-1)^{a\tilde{x}+\xi_x} \phi_{\mu,j}(x)]\\
&= \frac{1+\mu}{2}\cdot (-1)^{\xi_1}\cdot \phi_{\mu,j}(1) + \frac{1-\mu }{2}\cdot (-1)^{a+\xi_{-1}} \cdot \phi_{\mu,j}(-1).
\end{align*}
By plugging in we now obtain
\begin{align*}
\hat{g}_\mu^{(0)} (0) &= \frac{1+\mu}{2}\cdot (-1)^{\xi_1}\cdot 1 + \frac{1-\mu }{2}\cdot (-1)^{\xi_{-1}}\cdot 1,\\
\hat{g}_\mu^{(0)} (1) &= \frac{1+\mu}{2}\cdot (-1)^{\xi_1}\cdot \frac{1-\mu}{\sqrt{1-\mu^2}} + \frac{1-\mu }{2}\cdot (-1)^{\xi_{-1}}\cdot \frac{-1-\mu}{\sqrt{1-\mu^2}},\\
\hat{g}_\mu^{(1)} (0) &= \frac{1+\mu}{2}\cdot (-1)^{\xi_1}\cdot 1 + \frac{1-\mu }{2}\cdot (-1)^{1+\xi_{-1}}\cdot 1,\\
\hat{g}_\mu^{(1)} (1) &= \frac{1+\mu}{2}\cdot (-1)^{\xi_1}\cdot \frac{1-\mu}{\sqrt{1-\mu^2}} + \frac{1-\mu }{2}\cdot (-1)^{1+\xi_{-1}}\cdot \frac{-1-\mu}{\sqrt{1-\mu^2}}.
\end{align*}
So with probability $(\eta^1)^2 + (1-\eta^1)^2 = 1-2\eta^1(1-\eta^1)$, namely if $\xi_1 = \xi_{-1} = b\in\lbrace 0,1\rbrace$, we obtain
\begin{align*}
\hat{g}_\mu^{(0)} (0) &= (-1)^{b},\quad
\hat{g}_\mu^{(0)} (1)  = 0,\quad
\hat{g}_\mu^{(1)} (0)  = (-1)^{b}\mu,\quad
\hat{g}_\mu^{(1)} (1)  = (-1)^{b}\sqrt{1-\mu^2},
\end{align*}
and with probability $2\eta^1(1-\eta^1)$, namely if $\xi_1 = b \neq \xi_{-1}$, we obtain
\begin{align*}
\hat{g}_\mu^{(0)} (0) &= (-1)^{b}\mu,\quad
\hat{g}_\mu^{(0)} (1) = (-1)^{b}\sqrt{1-\mu^2},\quad
\hat{g}_\mu^{(1)} (0) = (-1)^{b},\quad
\hat{g}_\mu^{(1)} (1) = 0.
\end{align*}
Therefore we obtain: With probability $1-2\eta^1(1-\eta^1)$ the $\mu$-biased Fourier coefficients satisfy
\begin{align*}
\hat{g}_\mu^{(a)} (j) = 
\begin{cases}
(-1)^b (1-j),\hspace*{27.0mm} \textrm{for }a=0\\
(-1)^b ((1-j)\mu + j\sqrt{1-\mu^2})\ \quad \textrm{for }a=1
\end{cases},
\end{align*}
and with probability $2\eta^1(1-\eta^1)$ the $\mu$-biased Fourier coefficients satisfy
\begin{align*}
\hat{g}_\mu^{(a)} (j) = 
\begin{cases}
(-1)^b ((1-j)\mu + j\sqrt{1-\mu^2})\ \quad \textrm{for }a=0\\
(-1)^b (1-j),\hspace*{27.5mm} \textrm{for }a=1
\end{cases},
\end{align*}
which is exactly the claim for $n=1$. \hfill $\square$ \\

\noindent\textbf{Proof of Theorem \ref{ThmAmpBiasBVNoisy}:}\\
We want to prove that $\mathbb{P}[\textrm{Algorithm \ref{AlgNoisyAmpBV} does not output } a] \leq \delta,$ where the probability is w.r.t.~both the internal randomness of the algorithm and the random variables.\\
First observe that, due to $(i)$ in Theorem \ref{ThmBiasBVNoisyProb}, exactly the same reasoning as in the proof of Theorem \ref{ThmAmpBiasBV2} shows that the probability of observing $j_{n+1}=1$ in at most $k-1$ of the $m$ runs of Algorithm \ref{AlgNoisyBiasBV} (assuming $k\leq\frac{m}{2}$) is bounded by
\begin{align}
\mathbb{P}\left[\textrm{Bin}(m,\frac{1}{2})\geq m-k\right]\leq \exp\left(-\frac{2\left(\frac{m}{2} - k\right)^2}{m}\right). \label{eq:NoisyNoOfRuns}
\end{align}
We will now search for the number of observations of $j_{n+1}=1$ which is required to guarantee that the majority string is correct with high probability. Suppose we observe $j_{n+1}=1$ in $k$ runs of Algorithm \ref{AlgNoisyBiasBV}, $k\in 2 \IN$. Again we see that
\begin{align*}
\mathbb{P}[\exists 1\leq i\leq n: a_i \neq o_i]
\leq ~ &\mathbb{P}[\exists 1\leq i\leq n: a_i = 0 \neq o_i]\\
+ &\mathbb{P}[\exists 1\leq i\leq n: a_i = 1 \neq o_i].
\end{align*}
As ``false $1$'s'' can only appear in the case where our noise variables have an influence (compare Theorem \ref{ThmBiasBVNoisyProb}), we will first find a lower bound on $k$ which guarantees that the probability of the noise variable influence becoming relevant for at least $\frac{k}{5}$ runs is $\leq\frac{\delta}{4}$. Namely, we bound (again via Hoeffding)
\begin{align*}
\mathbb{P}[\textrm{Bin}(k, n\rho)\geq\frac{k}{5}]
&= \mathbb{P}[\textrm{Bin}(k, n\rho) - kn\rho\geq k(\frac{1}{5}-n\rho)]\\
&\leq \exp\left( -2k\left(\frac{1-5n\rho}{5}\right)^2\right).
\end{align*}
We now set this last expression $\leq \frac{\delta}{4}$ and rearrange the inequality to
\begin{align*}
k\geq \frac{25}{2(1-5n\rho)^2}\ln\left(\frac{4}{\delta}\right).
\end{align*}
Now we will find a lower bound on $k$ which guarantees that, if the noise variable influence is relevant in at most $\frac{k}{5}$ of the runs, among the remaining $\frac{4k}{5}$ runs with probability $\geq 1-\frac{\delta}{4}$ we make at most $\frac{k}{5}$ ``false $0$'' observations. To this end we bound (again via Hoeffding)
\begin{align*}
&~ \mathbb{P}[\textrm{Bin}(\frac{4k}{5}, n(1-c)^2)\geq\frac{k}{5}]\\
&= \mathbb{P}[\textrm{Bin}(\frac{4k}{5}, n(1-c)^2) - \frac{4kn(1-c)^2}{5}\geq \frac{k}{5} - \frac{4kn(1-c)^2}{5}]\\
&\leq \exp\left( - 2k(\frac{1}{5} - \frac{4n(1-c)^2}{5})^2\right).
\end{align*}
We now set this last expression $\leq \frac{\delta}{4}$ and rearrange the inequality to
\begin{align*}
k\geq \frac{25}{2(1-4n(1-c)^2)^2}\ln\left(\frac{4}{\delta}\right).
\end{align*}
Hence, by the union bound a sufficient condition for $\mathbb{P}[\exists 1\leq i\leq n: a_i\neq o_i]\leq\frac{\delta}{2}$ to hold is given by
\begin{align}
k\geq \frac{25}{2}\max\left\lbrace \frac{1}{(1-5n\rho)^2},\frac{1}{(1-4n(1-c)^2)^2}\right\rbrace\ln\left(\frac{4}{\delta}\right). \label{eq:NoisyNoOf1s}
\end{align}
Combining equations \eqref{eq:NoisyNoOf1s} and \eqref{eq:NoisyNoOfRuns} we now require
\begin{align*}
\exp\left(-\frac{2\left(\frac{25}{2}\max\left\lbrace \frac{1}{(1-5n\rho)^2},\frac{1}{(1-4n(1-c)^2)^2}\right\rbrace\ln\left(\frac{4}{\delta}\right) - \frac{m}{2}\right)^2}{m}\right)
\overset{!}{\leq}\frac{\delta}{4}.
\end{align*}
Rearranging gives the sufficient condition
\begin{align*}
m \geq 25\max\left\lbrace \frac{1}{(1-5n\rho)^2},\frac{1}{(1-4n(1-c)^2)^2}\right\rbrace\ln\left(\frac{4}{\delta}\right).
\end{align*}
This proves the claim of the theorem thanks to the union bound. \hfill $\square$\\

\noindent\textbf{Proof of Corollary \ref{CrlApproxBiasQuFourierTrafo}:}\\
According to the Lemma \ref{LmmApproxProdUnitaries} it holds that
\begin{align*}
&~ \norm{H_\mu - H_{\tilde{\mu}}}\\
&\leq\sum\limits_{i=1}^n \norm{\mathds{1}\otimes\ldots\otimes\mathds{1}\otimes H_{\mu_i}\otimes\mathds{1}\otimes\ldots\otimes\mathds{1} - \mathds{1}\otimes\ldots\otimes\mathds{1}\otimes H_{\tilde{\mu}_i}\otimes\mathds{1}\otimes\ldots\otimes\mathds{1}}\\
&= \sum\limits_{i=1}^n \norm{H_{\mu_i} - H_{\tilde{\mu}_i}}.
\end{align*}
Thus it suffices to bound the operator norm of the difference of the $1$-qubit biased quantum Fourier transforms. So let $|\varphi\rangle=\sum\limits_{x\in\lbrace -1,1\rbrace} \alpha_x|x\rangle$ be a qubit state. Then
\begin{align*}
(H_{\mu_j} - H_{\tilde{\mu}_j})|\varphi\rangle 
= \sum\limits_{x\in\lbrace -1,1\rbrace} \sum\limits_{j\in\lbrace 0,1\rbrace} \left(\sqrt{D_{\mu_i}(x)}\phi_{\mu_i,j}(x) - \sqrt{D_{\tilde{\mu}_i}(x)}\phi_{\tilde{\mu}_i,j}(x)\right)\alpha_x|j\rangle.
\end{align*}
We have to bound the (Euclidean) norm of this vector. To achieve this we will bound (for arbitrary $x\in\lbrace -1,1\rbrace$ and $j\in\lbrace 0,1\rbrace$) the expression
\begin{align*}
\left\lvert \sqrt{D_{\mu_i}(x)}\phi_{\mu_i,j}(x) - \sqrt{D_{\tilde{\mu}_i}(x)}\phi_{\tilde{\mu}_i,j}(x)\right\rvert^2.
\end{align*}
This is done by direct computation using $1-\mu_i^2 \geq 1-(1-c)^2\geq c^2$, $1-\tilde{\mu}_i^2\geq c^2$ and $|\mu_i-\tilde{\mu}_i|\leq\varepsilon$ as follows:
\begin{align*}
&\left\lvert\sqrt{D_{\mu_i}(x)}\phi_{\mu_i,j}(x) - \sqrt{D_{\tilde{\mu}_i}(x)}\phi_{\tilde{\mu}_i,j}(x)\right\rvert\\
&= \left\lvert \frac{(x_i-\mu_i)\sqrt{1-\tilde{\mu}_i^2}\sqrt{D_{\mu_i}(x)} - (x_i-\tilde{\mu}_i)\sqrt{1-\mu_i^2}\sqrt{D_{\tilde{\mu}_i}(x)}}{\sqrt{1-\tilde{\mu}_i^2}\sqrt{1-\mu_i^2}}\right\rvert\\
&\leq \frac{1}{c^2}\left\lvert(x_i-\mu_i)\sqrt{1-\tilde{\mu}_i^2}\sqrt{D_{\mu_i}(x)} - (x_i-\tilde{\mu}_i)\sqrt{1-\mu_i^2}\sqrt{D_{\tilde{\mu}_i}(x)}\right\rvert\\
&=\frac{1}{c^2}\left\lvert (x_i-\mu_i)\left(\sqrt{1-\tilde{\mu}_i^2}\sqrt{D_{\mu_i}(x)}-\sqrt{1-\mu_i^2}\sqrt{D_{\tilde{\mu}_i}(x)}\right) + (\tilde{\mu_i}-\mu_i)\sqrt{1-\mu_i^2}\sqrt{D_{\tilde{\mu}_i}(x)}\right\rvert\\
&\leq \frac{1}{c^2}\left(\left\lvert (x_i-\mu_i)\left(\sqrt{1-\tilde{\mu}_i^2}\sqrt{D_{\mu_i}(x)}-\sqrt{1-\mu_i^2}\sqrt{D_{\tilde{\mu}_i}(x)}\right)\right\rvert + \left\lvert (\tilde{\mu_i}-\mu_i)\sqrt{1-\mu_i^2}\sqrt{D_{\tilde{\mu}_i}(x)}\right\rvert\right)\\
&\leq \frac{1}{c^2}\left( (2-c)\left\lvert\sqrt{1-\tilde{\mu}_i^2}\sqrt{D_{\mu_i}(x)}-\sqrt{1-\mu_i^2}\sqrt{D_{\tilde{\mu}_i}(x)}\right\rvert + \varepsilon\right)\\
&\leq \frac{1}{c^2}\left( (2-c)\left(\left\lvert \sqrt{D_{\mu_i}(x)} - \sqrt{D_{\tilde{\mu}_i}(x)}\right\rvert + \left\lvert \sqrt{1-\mu_i^2} - \sqrt{1-\tilde{\mu}_i^2}\right\rvert\right)+ \varepsilon\right).
\end{align*}
Now note that
\begin{align*}
\left\lvert \left(\sqrt{D_{\mu_i}(x)} - \sqrt{D_{\tilde{\mu}_i}(x)}\right)\left(\sqrt{D_{\mu_i}(x)} +\sqrt{D_{\tilde{\mu}_i}(x)}\right) \right\rvert 
&= \left\lvert D_{\mu_i}(x) - D_{\tilde{\mu}_i}(x)\right\rvert\\
&= \left\lvert \frac{1+\tilde{x}_i\mu_i}{2} - \frac{1+\tilde{x}_i\tilde{\mu}_i}{2}\right\rvert\\
&= \frac{1}{2}|\mu_i-\tilde{\mu}_i|,
\end{align*}
which implies
\begin{align*}
\left\lvert\sqrt{D_{\mu_i}(x)} - \sqrt{D_{\tilde{\mu}_i}(x)}\right\rvert 
&= \left\lvert\frac{\mu_i-\tilde{\mu}_i}{2\left(\sqrt{D_{\mu_i}(x)} +\sqrt{D_{\tilde{\mu}_i}(x)}\right)}\right\rvert\\
&\leq \frac{\varepsilon}{2}\frac{1}{2\sqrt{\frac{c}{2}}}\\
&=\frac{\varepsilon}{2\sqrt{2c}},
\end{align*}
and that moreover
\begin{align*}
\left\lvert \left(\sqrt{1-\mu_i^2} - \sqrt{1-\tilde{\mu}_i^2}\right)\left(\sqrt{1-\mu_i^2} + \sqrt{1-\tilde{\mu}_i^2}\right)\right\rvert
&= \left\lvert 1-\mu_i^2 - (1-\tilde{\mu}_i^2)\right\rvert\\
&= \left\lvert \mu_i^2 - \tilde{\mu}_i^2\right\rvert,
\end{align*}
which in turn implies
\begin{align*}
\left\lvert\sqrt{1-\mu_i^2} - \sqrt{1-\tilde{\mu}_i^2}\right\rvert
&= \left\lvert\frac{\mu_i^2 - \tilde{\mu}_i^2}{\sqrt{1-\mu_i^2} + \sqrt{1-\tilde{\mu}_i^2}}\right\rvert\\
&\leq \frac{|\mu_i+\tilde{\mu}_i|\cdot |\mu_i-\tilde{\mu}_i|}{2\sqrt{1-(1-c)^2}}\\
&\leq \frac{2\varepsilon}{2\sqrt{2c-c^2}}\\
&\leq \frac{\varepsilon}{\sqrt{2}c}.
\end{align*}
Hence, we obtain
\begin{align*}
\left\lvert\sqrt{D_{\mu_i}(x)}\phi_{\mu_i,j}(x) - \sqrt{D_{\tilde{\mu}_i}(x)}\phi_{\tilde{\mu}_i,j}(x)\right\rvert
\leq \frac{1}{c^2}\left( (2-c)\left(\frac{\varepsilon}{2\sqrt{2c}} + \frac{\varepsilon}{\sqrt{2}c}\right) + \varepsilon\right)
\leq \gamma\varepsilon,
\end{align*}
where we defined $\gamma := \frac{1}{c^2}\left( (2-c)\frac{3}{2\sqrt{2}c} + 1\right)$. This now implies
\begin{align*}
\norm{(H_{\mu_j} - H_{\tilde{\mu}_j})|\varphi\rangle}_2
&\leq \sum\limits_{x\in\lbrace -1,1\rbrace} \sum\limits_{j\in\lbrace 0,1\rbrace} \norm{\left(\sqrt{D_{\mu_i}(x)}\phi_{\mu_i,j}(x) - \sqrt{D_{\tilde{\mu}_i}(x)}\phi_{\tilde{\mu}_i,j}(x)\right)\alpha_x|j\rangle}_2\\
&\leq \gamma\varepsilon\sum\limits_{x\in\lbrace -1,1\rbrace} \sum\limits_{j\in\lbrace 0,1\rbrace} |\alpha_x|\\
&= 2\gamma\varepsilon\sum\limits_{x\in\lbrace -1,1\rbrace}|\alpha_x|\\
&\leq 2\sqrt{2}\gamma\varepsilon.
\end{align*}
Finally, we get
\begin{align*}
\norm{H_\mu - H_{\tilde{\mu}}} \leq \sum\limits_{i=1}^n \norm{H_{\mu_i} - H_{\tilde{\mu}_i}} \leq 2\sqrt{2} n\gamma\varepsilon,
\end{align*}
as claimed.\hfill $\square$

\end{document}